\documentclass[a4paper,onecolumn,11pt,unpublished]{quantumarticle}
\pdfoutput=1

\usepackage[english]{babel}
\usepackage[T1]{fontenc}
\usepackage{amsmath}
\usepackage{amssymb}
\usepackage{graphicx}
\usepackage{microtype}
\usepackage{booktabs}
\usepackage{array}
\usepackage{xspace}

\usepackage{tikz}
\usetikzlibrary{arrows.meta,positioning,fit,calc,shapes.geometric,backgrounds}
\usepackage{quantikz}

\usepackage{listings}
\usepackage{xcolor}

\usepackage[numbers,sort&compress]{natbib}
\usepackage{hyperref}

\definecolor{pykw}{RGB}{0,63,136}
\definecolor{pystr}{RGB}{148,54,52}
\definecolor{pycom}{RGB}{100,100,100}
\definecolor{pybkw}{RGB}{98,84,148}
\definecolor{codebg}{RGB}{248,248,245}
\definecolor{codeframe}{RGB}{176,176,176}
\definecolor{irkw}{RGB}{0,63,136}

\definecolor{figblue}{RGB}{42,88,151}
\definecolor{figteal}{RGB}{0,118,110}
\definecolor{figamber}{RGB}{176,110,8}
\definecolor{figslate}{RGB}{92,101,115}

\lstdefinestyle{pyqbase}{
  basicstyle=\ttfamily\footnotesize,
  keywordstyle=\color{pykw}\bfseries,
  keywordstyle=[2]\color{pybkw},
  stringstyle=\color{pystr},
  commentstyle=\color{pycom}\itshape,
  numberstyle=\tiny\color{pycom},
  backgroundcolor=\color{codebg},
  frame=single,
  rulecolor=\color{codeframe},
  framerule=0.4pt,
  xleftmargin=1.2em,
  xrightmargin=0.6em,
  framesep=0.6em,
  breaklines=true,
  breakatwhitespace=true,
  postbreak=\mbox{\textcolor{pycom}{$\hookrightarrow$}\space},
  showstringspaces=false,
  showtabs=false,
  tabsize=4,
  captionpos=b,
  aboveskip=1.1em,
  belowskip=1.1em,
}

\lstdefinestyle{pyqpython}{
  style=pyqbase,
  language=Python,
  morekeywords=[2]{Protocol, runtime_checkable, isinstance, assert, self, None, True, False, print, with, as, partial},
  morestring=[b]{"""},
}

\lstdefinestyle{pyqjson}{
  style=pyqbase,
  language=,
  morekeywords={Program, Module, Register, RegType, Span, Ref, QRAM, Resource, Primitive, Load, Call, Repeat, Control, Adjoint, bits, uint, sint, rational, entry, modules, body, registers, resources, attributes, locals, name, count, parts, kind, width, start, tag, op, args, type},
  keywordstyle=\color{pybkw},
}

\lstdefinestyle{pyqir}{
  style=pyqbase,
  language=,
  morekeywords={DEF, ENDDEF, QRAMDECL, QINIT, CREG, H, X, Y, Z, S, T, RX, RY, RZ, U1, CNOT, SWAP, controlled_by, DAGGER, ENDDAGGER},
}

\newcommand{\tabhead}[1]{\textbf{#1}}
\newcommand{\ok}{\ensuremath{\checkmark}}
\newcommand{\noent}{\textendash{}}

\newcommand{\lvlI}{\ensuremath{\mathrm{I}}}
\newcommand{\lvlC}{\ensuremath{\mathrm{C}}}
\newcommand{\lvlE}{\ensuremath{\mathrm{E}}}
\newcommand{\lvlP}{\ensuremath{\mathrm{P}}}
\newcommand{\oracq}{oracq\xspace}

\newcommand{\secref}[1]{Section~\ref{#1}}
\newcommand{\figref}[1]{Fig.~\ref{#1}}

\newcommand{\casql}{Laboratory of Quantum Information, University of Science and Technology of China, Hefei, Anhui, 230026, China}
\newcommand{\casex}{Anhui Province Key Laboratory of Quantum Network, University of Science and Technology of China, Hefei 230026, China}
\newcommand{\aihf}{Institute of Artificial Intelligence, Hefei Comprehensive National Science Center, Hefei, Anhui, 230088, China}
\newcommand{\origin}{Origin Quantum Computing Technology (Hefei) Co., Ltd., Hefei, Anhui, 230026, China}

\title{oracq: A Modular Framework for Implementing and Analyzing Quantum Scientific Computing Algorithms}

\author{Zhao-Yun Chen}
\orcid{0000-0002-5181-160X}
\email{chenzhaoyun@iai.ustc.edu.cn}
\affiliation{\aihf}

\author{Ming-Yang Tan}
\affiliation{\casql}

\author{Guo-Ping Guo}
\orcid{0000-0002-2179-9507}
\affiliation{\casql}
\affiliation{\casex}
\affiliation{\aihf}
\affiliation{\origin}

\begin{document}

\begin{abstract}
Quantum algorithms for scientific computing are specified against access
models such as block-encodings, sparse oracles, state preparations, and
quantum random-access memories.
Their concrete realization is deliberately left open, yet it is a
research variable that changes both the numerical behavior and the
resource cost of the program, so every reported conclusion is
attributable to one realization.
In this paper, we present \oracq{}, a Python framework that makes this
choice a managed and checkable decision.
Our claim is linguistic rather than infrastructural.
Algorithms are stored as persistent register-level programs whose oracle
slots are declared and typed but unimplemented.
A separate binding pass links compatible realizations, so one stored
algorithm can be re-bound to a gate network, a QRAM-backed table, or
compiled reversible arithmetic and compared across them.
Typed QRAM resources, a compiler from pure Python numerical formulas to
reversible fixed-point circuits, and block-encoding algebra with
qubitization and QSVT-style kernels supply the constructions.
Two end-to-end workflows exercise the framework, a quantum finite-volume
pipeline for the compressible Euler equations and a quantum
homotopy-analysis pipeline that lifts nonlinear PDEs to composable
linear input models.
The verification methodology separates implementation error from method
error, and logical resource accounts cover programs before and after
binding.
The framework is openly available at
\url{https://github.com/IAI-USTC-Quantum/oracq}.
\end{abstract}

\maketitle

\section{Introduction}
\label{sec:introduction}

Quantum algorithms for scientific computing are specified relative to
\emph{access models}.
Linear-system solvers
\cite{harrow2009hhl,childs2017qls,costa2022discrete}, Hamiltonian
simulation \cite{berry2015taylor,low2019qubitization}, and the
differential-equation methods built on them
\cite{gilyen2019qsvt,martyn2021grand,an2026lchs,jin2024schrodingerization,liu2021pnas}
all take this form.
An access model may be a block-encoding of a matrix, sparse location
and entry oracles, a preparation of an input state, or a quantum
random-access memory (QRAM).
The access model is the algorithm's unit of abstraction, and its
concrete realization is deliberately left open.
Matrix entries may be loaded from a table or computed from physical
variables.
State preparation may use gates or a memory.
Finite-precision arithmetic can substantially change both the
numerical result and the resource requirements.
A theory paper fixes the access model and moves on.
An implementation must commit to a realization, and every numerical
and resource conclusion it reports is a conclusion about that
realization.

Our primary user is the quantum algorithm researcher who implements,
modifies, and compares these constructions.
For this user, the challenge is not only to generate a circuit.
It is to preserve the mathematical algorithm while making its access
assumptions concrete, to compose subroutines with compatible
interfaces, and to understand which numerical and resource conclusions
are justified by each realization.
These concerns run across the field rather than through any single
application.
Linear differential equations are addressed by LCHS
\cite{an2026lchs}, Schr\"odingerization
\cite{jin2024schrodingerization,hu2024schrocircuits}, and
contour-based matrix decomposition \cite{wang2026cbmd}.
Nonlinear equations are addressed by Carleman lifting
\cite{liu2021pnas,liu2023carleman} and the quantum homotopy analysis
method (QHAM) \cite{xue2025qham}.
Fluid dynamics is addressed by the quantum finite-volume method (QFVM)
\cite{chen2022qfvm}, hybrid fluid simulation on superconducting
hardware \cite{chen2024fluid}, and end-to-end Navier--Stokes resource
studies \cite{zhuang2025navierstokes}.
Across these subfields, an implementation must connect physical data,
numerical formulas, linear-algebra kernels, and output recovery.
None of these connections should change silently when a component is
replaced.

Existing tools support parts of this workflow.
Embedded circuit languages and frameworks generate and lower quantum
programs, and algorithm-analysis platforms such as
Qualtran~\cite{harrigan2024qualtran} and
pyLIQTR~\cite{obenland2026pyliqtr} provide composable block-encoding
constructions with resource estimation.
None of them, however, supports the researcher's loop as a whole, and
\secref{sec:motivation} examines what a new programming language must
supply.

In this paper, we present \oracq{}, a modular Python framework
organized around this loop.
An embedded interface generates a register-level intermediate
representation (RIR).
A program is a graph of modules with explicit register interfaces,
structured repetition, control, and adjoint, and with no host-language
callbacks.
An oracle slot is an open declaration with a typed signature, declared
capabilities, and fixed mathematical conventions such as a
block-encoding normalization.
Open programs are persistent.
They can be validated, serialized, and restored without the original
Python generator.
A separate binding pass then links compatible implementations, such as
gate networks, QRAM-backed tables, or compiled arithmetic.
The pass captures the memory resources these implementations introduce
and reports what remains unresolved.

On this core, two layers make scientific algorithms expressible.
Python structural protocols and algorithm contracts connect
application objects to program generation without extending the IR.
A library of block-encoding algebra with qubitization and QSVT-style
kernels assembles the linear-algebra layers.
The data side of scientific access models rests on two complementary
mechanisms.
Typed QRAM resources with pointer-style array views move coherent data
at run time.
A compiler turns pure Python numerical formulas into reversible
fixed-point circuits.

The framework makes three contributions.
\begin{enumerate}
\item \textbf{A persistent separation between algorithms and oracle
realizations.} Open modules are program objects, not Python callbacks.
Checked partial binding preserves their interfaces, captures
implementation resources, and records remaining obligations.
The same artifact can be analyzed before and after its implementations
are selected.
\item \textbf{QRAM access and compiled reversible arithmetic.} Typed
QRAM resources move coherent data with XOR load semantics, restricted
classical writes, and per-bank query accounting.
A compiler turns pure Python numerical formulas into reversible
fixed-point circuits with explicit formats, domain and overflow
status, and workspace conventions.
The two mechanisms realize table-based and formula-based oracles
interchangeably under the binding checks of the first contribution.
\item \textbf{Composable scientific-computing interfaces.} Role views
and contracts connect block-encodings, state preparations, sparse
access, and solver generators, so algorithm libraries declare, check,
and compose their input models without extending the core
representation.
\end{enumerate}

Because realizations are replaceable by design, their evaluation is
part of the workflow rather than an afterthought.
QFVM and QHAM serve as end-to-end case studies exercising different
paths from scientific inputs to composed algorithms.
Compiled programs run against references constructed independently of
the tested code, so implementation error is measured separately from
method error.
A controlled study restores one stored open algorithm and binds it to
a gate table, a QRAM table, and compiled arithmetic, and the
realizations agree numerically while their logical resource accounts
differ.
Every numerical claim is instance-scoped, and the tested configuration,
reference, and acceptance criterion are recorded with each case.

The remainder of this section argues why the missing pieces are a new
programming language rather than better engineering
(\secref{sec:motivation}), and then surveys related frameworks and
algorithmic precedents in that light (\secref{sec:related}).
\secref{sec:requirements} then presents the framework and its
representation, covering the programming interface
(\secref{sec:language}), the register-level program representation
(\secref{sec:rir}), the oracle lifecycle with checked partial binding
(\secref{sec:oracle-lifecycle}), and the role, protocol, and contract
layer (\secref{sec:protocols}).
\secref{sec:primitives} develops the block-encoding constructions.
\secref{sec:arithmetic} treats the two data-side mechanisms, QRAM
resources and compiled reversible arithmetic, together with the
compiler that turns a Python formula into a reversible circuit.
\secref{sec:ode} assembles the scientific workflows QFVM
(\secref{sec:qfvm}) and QHAM (\secref{sec:qham-case}).
\secref{sec:verification} sets out how quantum scientific-computing
programs are verified and reports the validation and resource evidence.
\secref{sec:discussion} discusses the limits of the present claims.
Reproduction commands and cost conventions are collected in the
appendix.

\subsection{Why a new language, and not better engineering}
\label{sec:motivation}

It is fair to ask what a new quantum programming language adds in the era
of \emph{vibe coding}, when much of the code for a quantum program can be
generated on demand. Our answer is that the missing pieces are linguistic
rather than infrastructural, and that they are exactly the pieces that make
generated code trustworthy.

\emph{First, late binding of oracle realizations is not part of the
persisted program.} Almost every existing quantum programming language
resolves its instruction set at compile time. Each oracle is expanded into
a concrete circuit before the program exists as an object. The closest
mechanisms stop short of a shareable open artifact. Qualtran's bloqs carry
register signatures and can stay abstract for symbolic resource
counting~\cite{harrigan2024qualtran,qualtran2026interfaces}, but an
undecomposed bloq survives serialization only as a shallow
name-and-signature entry whose restoration resolves against a registry of
concrete classes, and composite bloqs are documented as immutable. What is
not offered is a program whose access slots are declared and typed but
deliberately unimplemented, stored as a shareable artifact, and closed
later under an interface and convention check. The same slot can be bound
repeatedly, each time to a different realization. The realization chosen at
compilation time remains the only one the program can ever have.

\emph{Second, register-level structure persists only inside compilers.}
Existing IRs are gate-level or pulse-level, and the algorithm is recovered,
if at all, from the circuit that implements it. Catalyst's MLIR dialects do
hold a structured hybrid representation of the classical--quantum
program~\cite{catalyst2026compiler}, but as a transient compiler artifact
lowered to LLVM/QIR and a binary, not as the distributable artifact itself.
What is missing is a representation that keeps the program's mathematical
structure as an explicit object that survives generation, serialization,
and restoration. That structure includes register interfaces, structured
repetition, control, and adjoint, and its open slots participate in
checked, partial binding.

\emph{Third, these two gaps make most published implementations either
empty or hard to verify.} The dominant failure mode in the current
literature is an algorithm specified up to a \emph{general} oracle and
therefore testable only in principle. A declared, substitutable access slot
inverts this. The upper-level program can be checked against a single
constructed, simple oracle, and its numerical behavior measured separately
from the behavior of the implementation. The same gap appears in
composition. A complex algorithm such as QHAM~\cite{xue2025qham} cannot
readily reuse the Schr\"odingerization~\cite{jin2024schrodingerization} or
LCHS~\cite{an2026lchs} subroutines as components, because each is
re-derived and re-implemented rather than linked.

A further, though secondary, gap is surface syntax. QRAM access and
arbitrary-precision reversible arithmetic are still hand-rolled idioms in
most languages, written out by every author who needs them. They are small
pieces of syntactic sugar, but they are missing parts of the languages that
scientific quantum programs are currently written in.

Our claim is therefore not one of engineering improvement but of
programming paradigm. Once algorithms are defined as programs over declared
access models, complex programs become programmatically composable and
their upper levels become verifiable against replaceable, checked
realizations. That is a property the language either provides or does not.

\subsection{Related frameworks and algorithmic precedents}
\label{sec:related}

We now assess existing systems against the three gaps identified above.
Qualtran~\cite{harrigan2024qualtran} and pyLIQTR
\cite{obenland2026pyliqtr} are the closest precedents for compositional
fault-tolerant algorithm implementation and analysis. Qualtran organizes
computations as named bloqs with register signatures, decomposition and
call-graph interfaces, resource analysis, simulation, and serialization.
A bloq can specify symbolic call counts without a complete decomposition
\cite{qualtran2026interfaces}.
Our emphasis is the complete sequence of restoring a stored open program,
checking replacement interfaces, capturing QRAM resources, and analyzing
the resulting scientific computation before and after partial linking.
pyLIQTR provides related block-encoding and QSP/QSVT circuit construction
and resource-analysis tools. These systems are direct comparators for
our component and analysis layers.
The BlockEncoding interface of Qrisp~\cite{petric2026qrispbe} is the most
recent direct comparator at the component layer. It covers block-encoding
construction from arrays, physics operators, and LCU decompositions, an
operator-overload algebra with tracked subnormalization (sums, products,
tensor products, adjoints), qubitization walks and Chebyshev polynomials,
matrix inversion by QET/QSVT/GQSVT, and simulator-traced gate, depth, and
qubit queries, with turnkey solvers (CKS, a QSVT linear-system solver)
assembled on top. The interface's objects are concrete Python callables
resolved at construction. There is no persisted open program, no
binding-time convention check, and no named runtime QRAM resource. The
documented resource queries are simulator traces of one execution rather
than a fault-tolerant ledger over a stored program.

Table~\ref{tab:framework-features} summarizes feature availability in
the documented framework interfaces. Partial support is distinguished
from a direct implementation. The footnotes clarify the few cases where
similar mechanisms serve different purposes.
The table concerns mechanisms rather than algorithms. The algorithm and
solver coverage of the major frameworks' standard libraries and of the
ecosystem libraries, \oracq{} included, is recorded separately in
Table~\ref{tab:ode-coverage}.

\begin{table}[t]
\centering
\caption{Feature availability in the closest algorithm-analysis frameworks.
\ok~= directly provided. $\triangle$~= partial or related support.
\noent~= no direct facility identified in the reviewed documentation.
The sources are the Qualtran
\cite{harrigan2024qualtran,qualtran2026interfaces,qualtran2026library},
pyLIQTR~\cite{obenland2026pyliqtr,pyliqtr2026features},
and Qrisp~\cite{petric2026qrispbe,seidel2023qrisp}
documentation.
Algorithm and solver coverage, standard libraries and ecosystem
libraries alike, is recorded in Table~\ref{tab:ode-coverage}. The
validation limits of \oracq{} are discussed in
\secref{sec:implementation}.}
\label{tab:framework-features}
\small
\setlength{\tabcolsep}{4pt}
\begin{tabular}{@{}>{\raggedright\arraybackslash}p{5.9cm}cccc@{}}
\toprule
Feature & Qualtran & pyLIQTR & Qrisp & \oracq{} \\
\midrule
Typed modular composition & \ok & \ok & \ok & \ok \\
Block-encoding components & \ok & \ok & \ok & \ok \\
QSP/QSVT circuit construction & \ok & \ok & \ok & \ok$^{a}$ \\
Structured program serialization & \ok & $\triangle^{b}$ & \noent & \ok \\
Deferred oracle realization with partial linking & $\triangle^{c}$ & $\triangle^{c}$ & \noent & \ok \\
Named runtime QRAM resources$^{d}$ & \noent & \noent & \noent & \ok \\
Python-formula-to-reversible-circuit compiler & \noent & \noent & \noent & \ok \\
QFVM/QHAM workflow components & \noent & \noent & \noent & \ok \\
Logical resource estimation & \ok & \ok & \ok & \ok \\
Surface-code physical cost models & \ok & \noent & \noent & \noent \\
\bottomrule
\end{tabular}
\par\smallskip
\begin{minipage}{0.96\linewidth}
\footnotesize
$^{a}$Circuit constructors are available. The \oracq{} bundled phase
synthesizer has a degree guard and numerical limits. Qrisp delegates phase
synthesis to internal solvers without a documented degree or precision
limit.
$^{b}$Cirq circuits and OpenQASM export, rather than a dedicated stored
program with open-slot linking.
$^{c}$Qualtran supports symbolic call counts without complete
decompositions. pyLIQTR supports abstract and alternative
block-encodings.
These are related to, but not the full persisted partial-linking workflow.
$^{d}$Runtime memory identities and separate data binding, not merely
QROM circuits. Writes in \oracq{} are restricted classical updates.
\end{minipage}
\end{table}

Earlier programming systems establish important ingredients.
Quipper supports hierarchical circuit construction
\cite{green2013quipper}. ProjectQ supplies Python circuit generation,
control and inverse contexts, and compiler engines
\cite{steiger2018projectq}. Q\# tracks controlled and adjoint
specializations~\cite{svore2018qsharp}. RIR similarly preserves structured
operations, while making declared access models and repeated partial
binding part of the stored artifact.
Their standard libraries also anchor the algorithmic baseline.
Q\# ships fixed-point numerics and estimation primitives, but no
block-encoding algebra, QSP machinery, or ODE/PDE solver families.
Qiskit~\cite{javadiabhari2024qiskit} retired the demonstration-level HHL
circuit of its former algorithms module and carries no QSVT or
block-encoding abstraction today, while the ProjectQ and Cirq
\cite{cirqdevelopers2020cirq} libraries center on NISQ circuit
primitives. Table~\ref{tab:ode-coverage} records these differences row
by row, together with the corresponding components of the \oracq{}
library.

Table~\ref{tab:ode-coverage} therefore scores two boundaries. The
upper panel counts each major framework's own standard library only.
The lower panel turns to the ecosystem and commercial libraries
introduced above.
Qualtran and pyLIQTR supply the block-encoding and QSP/QSVT components
of the analysis-library layer.
PennyLane ships subroutine templates (\texttt{QSVT}, \texttt{GQSP},
\texttt{Qubitization}, \texttt{FABLE}) and built-in phase-angle generation
(\texttt{poly\_to\_angles}, documented to validate the synthesized circuit
against the target polynomial at the $10^{-10}$
level)~\cite{bergholm2018pennylane,pennylane2026polytoangles}, rather than
packaged solver families.
Classiq's example library assembles QSVT matrix inversion, HHL, and
quantum differential-equation demonstrations over a closed synthesis
engine~\cite{classiq2026platform}.
UnitaryLab's algorithm library spans HHL, a QSVT linear-system solver,
several Hamiltonian-simulation kernels, and Schr\"odingerization-based
heat- and advection-equation
solvers~\cite{unitarylab2025platform}.
Under this ecosystem reading, several middle rows do find
implementations outside the standard libraries. The distinguishing
claim of \oracq{} is the language-level integration and the ODE/PDE
solver families.

\begin{table}[p]
  \centering
  \caption{Algorithm and solver maturity of \oracq{} against the major
  frameworks' standard libraries (upper panel) and against the
  quantum-scientific-computing ecosystem and commercial libraries
  (lower panel).  The upper panel counts each framework's \emph{own}
  standard library only. The lower panel scores the public libraries
  named in its columns by their documentation
  \cite{harrigan2024qualtran,qualtran2026library,obenland2026pyliqtr,pyliqtr2026features,bergholm2018pennylane,pennylane2026polytoangles,classiq2026platform,unitarylab2025platform,petric2026qrispbe},
  so that no side is credited or discounted by the choice of boundary.
  Each cell is graded by maturity. \lvlI~= declared interface.
  \lvlC~= circuit generation. \lvlE~= closed execution (simulate or run
  end-to-end). \lvlP~= specified numerical precision reached with
  documented validation. \noent~= absent from the reviewed documentation.
  Where the documentation is ambiguous between two levels, the lower one
  is taken. The \oracq{} column is graded against the witnesses of
  \secref{sec:implementation}. Its single \lvlE{} cell is a prototype
  kernel whose correctness is marked pending.}
  \label{tab:ode-coverage}
  \footnotesize
  \setlength{\tabcolsep}{4pt}\begin{tabular}{@{}p{5.4cm}ccccc@{}}
    \toprule
    \tabhead{Capability} & \textbf{Q\#} & \textbf{ProjectQ} & \textbf{Qiskit} & \textbf{Cirq} & \textbf{\oracq} \\
    \midrule
    Query/search/amplification & \lvlE & \lvlE & \lvlE & \lvlE & \lvlP \\
    Phase/amplitude estimation & \lvlE & \lvlE & \lvlE & \lvlC & \lvlP \\
    Hamiltonian simulation (Trotter) & \lvlE & \lvlC & \lvlE & \lvlC & \lvlP \\
    Hamiltonian simulation (LCU/Taylor, QSVT) & \noent & \noent & \lvlC & \noent & \lvlP \\
    Block-encoding algebra & \noent & \noent & \lvlC & \noent & \lvlP \\
    QSP phase synthesis (self-verifying) & \noent & \noent & \noent & \noent & \lvlP \\
    Quantum linear systems (QLSS) & \noent & \noent & \lvlC$^{a}$ & \noent & \lvlP \\
    QRAM/QROM input models (typed resource) & \noent & \noent & \lvlC & \noent & \lvlP \\
    Reversible fixed-point arithmetic & \lvlE & \noent & \noent & \noent & \lvlP \\
    Math-function compiler (Python $\rightarrow$ reversible) & \noent & \noent & \noent & \noent & \lvlP \\
    ODE (LCHS) & \noent & \noent & \noent & \noent & \lvlP \\
    ODE (Schr\"odingerization) & \noent & \noent & \noent & \noent & \lvlP \\
    ODE (contour decomposition, CBMD) & \noent & \noent & \noent & \noent & \lvlP \\
    ODE (Carleman, nonlinear) & \noent & \noent & \noent & \noent & \lvlP \\
    ODE (history systems via QLSS) & \noent & \noent & \noent & \noent & \lvlE \\
    PDE (heat / Fokker--Planck input models) & \noent & \noent & \noent & \noent & \lvlP \\
    PDE (symbolic nonlinear PDEs, QHAM) & \noent & \noent & \noent & \noent & \lvlP \\
    PDE (Euler equations, QFVM/Roe) & \noent & \noent & \noent & \noent & \lvlP \\
    \bottomrule
    \multicolumn{6}{@{}p{14.0cm}@{}}{\scriptsize $^a$A demonstration-level HHL circuit shipped historically in Qiskit's (since-retired) algorithms module. It is not part of the current library.}
  \end{tabular}

  \par\smallskip
  \setlength{\tabcolsep}{2.5pt}\begin{tabular}{@{}p{2.9cm}ccccccc@{}}
    \toprule
    \tabhead{Capability} & \textbf{Qualtran} & \textbf{pyLIQTR} & \textbf{PennyLane} & \textbf{Classiq} & \textbf{UnitaryLab} & \textbf{Qrisp} & \textbf{\oracq} \\
    \midrule
    Hamiltonian simulation (LCU/Taylor, QSVT) & \lvlC & \lvlC$^{b}$ & \lvlE$^{b}$ & \lvlE$^{g}$ & \lvlE$^{h}$ & \lvlE$^{i}$ & \lvlP \\
    Block-encoding algebra & \lvlC & \lvlC & \lvlE$^{b}$ & \lvlE & \lvlC$^{c}$ & \lvlP$^{i}$ & \lvlP \\
    QSP/QSVT phase machinery & \lvlC & \lvlC & \lvlP$^{b}$ & \lvlP$^{g}$ & \lvlE & \lvlP$^{i}$ & \lvlP$^{d}$ \\
    Quantum linear systems (QLSS) & \lvlI$^{e}$ & \noent & \lvlE$^{b}$ & \lvlP$^{g}$ & \lvlE$^{h}$ & \lvlP$^{i}$ & \lvlP \\
    ODE/PDE solver families & \noent & \noent$^{f}$ & \noent & \lvlE$^{g}$ & \lvlE$^{h}$ & \noent & \lvlP \\
    \bottomrule
    \multicolumn{8}{@{}p{14.6cm}@{}}{\scriptsize $^b$PennyLane's \texttt{QSVT}, \texttt{GQSP}, \texttt{Qubitization}, \texttt{FABLE}, and \texttt{BlockEncode} templates run on its simulators only. \texttt{poly\_to\_angles} adds phase generation documented to validate at the $10^{-10}$ level. Packaged solver families are not shipped (the QLSS entry rests on validated demonstrations).  $^c$Block encodings underlie UnitaryLab's QSVT linear solver. The Schr\"odingerization path documents its block-encoding backend as pending.  $^d$Computed phases are checked against the polynomial identity during validation, and the synthesizer is a replaceable, contract-checked component.  $^e$Block-encoding and QSP bloqs exist, but no assembled solver.  $^f$The \texttt{clam} layer integrates ODEs classically. No quantum solver families ship with it.  $^g$QSVT matrix inversion, HHL, and quantum differential-equation examples in the open example library over a closed synthesis engine.  $^h$Solvers execute over numpy/torch backends. The Schr\"odingerization \texttt{block} option is documented to fall back to classical computation.  $^i$Qrisp's \texttt{inv()}/CKS ship with a documented numeric comparison against the classical solution. Its resource queries are simulator traces of one execution.}
  \end{tabular}
\end{table}

The mathematical methods behind the lower half of
Table~\ref{tab:ode-coverage} are complementary to the framework.
Today each exists as research code accompanying its publication, and
none is a library component of any framework's standard library.
LCHS~\cite{an2026lchs}, Schr\"odingerization
\cite{jin2024schrodingerization,hu2024schrocircuits}, contour-based matrix
decomposition (CBMD)~\cite{wang2026cbmd}, and Carleman lifting
\cite{liu2021pnas,liu2023carleman} supply distinct reductions to linear
algebra or Hamiltonian simulation. QHAM~\cite{xue2025qham} supplies the
quantum-compatible linearization used in our nonlinear PDE example.
Finite-volume quantum CFD~\cite{chen2022qfvm}, hybrid fluid simulation on
superconducting hardware~\cite{chen2024fluid}, and end-to-end
Navier--Stokes resource studies~\cite{zhuang2025navierstokes} motivate the
need to treat input, arithmetic, output, and resource assumptions together.
Our interfaces support implementations of these constructions. Their
mathematical assumptions do not become interchangeable merely because
their software interfaces are compatible.

Finally, logical accounting must be distinguished from physical
fault-tolerant estimation. Application-level resource frameworks
\cite{beverland2022assessing,harrigan2024qualtran} incorporate additional
hardware and error-correction assumptions. Here we report logical gates,
workspace, and QRAM accesses from concrete lowering recipes.

\section{Framework and deferred oracle realization}
\label{sec:requirements}

The workflow separates the mathematical construction, its access-model
realization, and its execution environment.
Python is used to generate an algorithm against declared interfaces.
RIR stores the resulting module graph, including unimplemented slots.
Binding chooses compatible implementations. Export or execution then
consumes a closed program. Numerical validation and resource analysis
refer to these explicit artifacts rather than to hidden host callbacks.

\subsection{Programming with registers and modules}
\label{sec:language}

Before describing RIR, we show how a researcher writes a small program.
The Python interface constructs quantum operations. Running the Python
generator does not itself execute a quantum circuit. Listing~\ref{lst:bell}
prepares a Bell state using one named two-bit register. Index~0 denotes
its least significant bit. Indexing a register creates a view. The
\texttt{h} and \texttt{xor} calls emit operations on those views.

\begin{lstlisting}[style=pyqpython, caption={A complete first program. The dependency-free reference executor returns the two nonzero output amplitudes. No native backend is needed to construct the program.}, label={lst:bell}]
from oracq import Bits, Builder, simulate

b = Builder("bell_pair", {"pair": Bits(2)})
b.h(b["pair"][0])
b.xor(b["pair"][0], b["pair"][1])
program = b.finish().program()
state = simulate(program).amplitudes
print(state)
# {(0,): (0.7071067811865475+0j), (3,): (0.7071067811865475+0j)}
\end{lstlisting}

\texttt{finish()} returns an \texttt{Operation}, consisting of the entry
module and its dependencies. \texttt{program()} collects and validates the
module graph. The program remains data until a reference executor or
backend consumes it. Its public register name and width are retained.

The same program can defer the entangling operation. In
Listing~\ref{lst:open-bell}, \texttt{declare} fixes a one-bit control and
target interface without supplying a body. The caller records a
\texttt{Call}. The saved program remains open. A compatible implementation
then closes that slot without regenerating the caller.

\begin{lstlisting}[style=pyqpython, caption={Refactoring the Bell program into an open module and binding its implementation. This continues Listing~\ref{lst:bell}. \texttt{dumps} and \texttt{loads} preserve the open call and register views.}, label={lst:open-bell}]
from oracq import bind, declare, dumps, loads

slot = declare("Entangle", {"control": Bits(1), "target": Bits(1)})
b = Builder("open_bell", {"pair": Bits(2)})
b.h(b["pair"][0])
b.call(slot, control=b["pair"][0], target=b["pair"][1])
opened = loads(dumps(b.finish().program()))

impl = Builder("CNOT", {"control": Bits(1), "target": Bits(1)})
impl.xor(impl["control"], impl["target"])
closed = bind(opened, {"Entangle": impl.finish()})
print(simulate(closed).amplitudes == state)
# True
\end{lstlisting}

For static repetition the author writes \texttt{with b.repeat(n):} around
emitted operations. \texttt{with b.control(view):} and
\texttt{with b.adjoint():} similarly create structured control and
inversion blocks. These are generation-time Python scopes that become
single RIR nodes, rather than Python callbacks stored in the program.
Application-level access methods such as \texttt{block\_encoding()} then
provide typed views of such operations to algorithm generators, as shown
in Section~\ref{sec:protocols}.

Three requirements guide this separation.
First, an incomplete realization must still be a valid, shareable
algorithm artifact. All callees have declared interfaces even when some
bodies are absent.
Second, composition must preserve the information needed by later
stages. These include register views, resource identities, controlled
and adjoint calls, and repetition counts.
Third, replacing an implementation must not silently change an
algorithm's mathematical input conventions. A fixed block-encoding scale
or a finite-precision layout used during generation cannot simply be
relabelled at linking time.
These requirements motivate the distinction between RIR modules, Python
role views, and algorithm contracts developed below.

The core package does not depend on native quantum backends. Native
simulators are optional and imported only at execution entry.
The Python reference executor supports small-instance checks. The
PySparQ adapter, the native PySparQ RIR interpreter, and the
OriginIR-ext/UnifiedQuantum path provide additional execution routes.
Backend acceleration is separate from the logical gate-cost model.

Table~\ref{tab:feature-summary} summarizes the features that support this
workflow. It describes the framework's interfaces and mechanisms.
Implementation coverage against other frameworks' standard libraries is
reported in \secref{sec:related} (Table~\ref{tab:ode-coverage}), and the
extent of validation in Table~\ref{tab:verify-matrix}.

\begin{table}[t]
\centering
\caption{Core features of \oracq{}. Each feature has a distinct role in
the scientific-algorithm workflow.}
\label{tab:feature-summary}
\small
\begin{tabular}{@{}>{\raggedright\arraybackslash}p{3.0cm}>{\raggedright\arraybackslash}p{8.2cm}p{1.6cm}@{}}
\toprule
Feature & Supported mechanism & Section \\
\midrule
Persistent open programs & Store and restore declared oracle slots together with their callers, independently of the original Python generator. & \ref{sec:rir} \\
Checked partial binding & Replace compatible implementations in stages, propagate resource captures, and retain unresolved obligations in a binding report. & \ref{sec:oracle-lifecycle} \\
Role-based inputs & Application objects supply access views through structural protocols. Contracts check and reuse the resolved inputs. & \ref{sec:protocols} \\
Structured composition & Preserve register views, calls, static repetition, control, and adjoint through generation and serialization. & \ref{sec:rir} \\
Explicit data resources & Treat QRAM banks and runtime tables separately from program text, with XOR reads and restricted classical writes. & \ref{sec:qram} \\
Reversible numerical oracles & Compile supported Python formulas to fixed-point compute/update/uncompute modules with domain and overflow status. & \ref{sec:mathfunc} \\
Compositional cost analysis & Separate known costs from unresolved oracle terms. Report logical gates, workspace, and accesses by memory resource. & \ref{sec:resources-summary} \\
Backend separation & Generate without native dependencies. Export or execute closed programs through distinct backend paths. & \ref{sec:requirements} \\
\bottomrule
\end{tabular}
\end{table}

\subsection{A register-level program representation}
\label{sec:rir}

At the semantic level, a program is a pair
$\langle\text{entry},\text{modules}\rangle$.
A module contains an ordered interface of quantum registers and QRAM
resources, optional private workspace, attributes, and an instruction
body. A body may be absent, representing an open declaration.
An empty body instead denotes an implemented identity. It is not an
unresolved oracle.

Registers retain their names, kinds, and widths. The storage kinds are
bits, unsigned and signed integers, and fixed-point-compatible rational
words. A register is limited to 64 bits, but a program can contain many
registers. A view is an ordered collection of spans
$(\text{register},\text{start},\text{width})$, allowing slices and
concatenations without emitting quantum operations.
Backend lowering, not the algorithm generator, assigns physical qubit
indices. Thus the public data layout remains inspectable through
composition and serialization.
Table~\ref{tab:rir-nodes} gives the instruction boundary.

\begin{table}[t]
\centering
\caption{The RIR instruction boundary. Composite algorithms are module
graphs over these nodes. Domain-specific access roles do not introduce
new instruction kinds.}
\label{tab:rir-nodes}
\small
\begin{tabular}{@{}lp{11.5cm}@{}}
\toprule
Node & Meaning \\
\midrule
Primitive & Register-level gates and reversible bit operations. Angles and integer parameters are fixed during generation. \\
Load & XOR access to a named QRAM resource. Address and data remain explicit register views. \\
Store & Restricted update of classical memory at definite address and data values. This is not a coherent superposition write. \\
Call & Invocation of a named module with ordered register and resource arguments. \\
Repeat & A static repetition count and one copy of its body. \\
Control & Coherent conditioning on protected register views. \\
Adjoint & Symbolic inversion of a body supporting adjoint execution. \\
\bottomrule
\end{tabular}
\end{table}

An embedded builder emits these nodes, while Python loops and expressions
can select concrete layouts and parameters before generation.
RIR itself contains no opaque Python functions, arbitrary symbolic widths,
measurement, reset, or runtime classical feedback. Coherent outputs can
therefore be consumed by later quantum subroutines. Measurement and
post-processing are separate host-side readout plans.
The restricted Store effect is an explicit exception to the unitary
fragment, described in Section~\ref{sec:qram}.

\paragraph{Composition and workspace.}
Validation checks register kinds and widths, nonoverlapping operands,
protected controls, call targets, and acyclic module dependencies.
Controlled and adjoint capabilities are inferred conservatively from the
reachable module graph. A context requesting an unavailable capability is
rejected before execution.
Private locals contribute to the live workspace requirements of a call
without becoming part of its mathematical public interface.
Zero-input and clean-workspace requirements are algorithm promises.
Structural validation does not prove that arbitrary circuits uncompute
their workspace.

\paragraph{Persistence and lowering.}
Canonical JSON stores the graph with deterministic ordering and is
checked against a schema. Decoding rejects unknown fields and tags,
duplicate keys, and non-finite numbers. Serialization retains open
declarations and does not replicate a Repeat body according to its count.
Restoring an artifact therefore does not require rerunning the original
algorithm generator.

OriginIR-ext export preserves module definitions and calls. Repetition
can be represented by shared helper definitions rather than repeated
text. Register operations are lowered only at the backend boundary.
Reference interpretation and native execution may expand operations
lazily under explicit budgets. These execution choices do not change the
stored RIR. Complete machine-readable schemas and backend grammars belong
to the reference documentation.

\subsection{Oracle lifecycle and checked partial binding}
\label{sec:oracles}
\label{sec:oracle-lifecycle}

An open declaration fixes an access interface while leaving its body
unspecified. Its register layout, declared capabilities, resource
parameters, and relevant mathematical conventions are already concrete.
For example, a block-encoding slot specifies target and signal widths and
a normalization $\alpha$. A database slot specifies address and data
widths. Access-role names are extensible library conventions, not a
closed enumeration of language types.

\begin{figure}[t]
  \centering
  \begin{tikzpicture}[
      font=\small,
      state/.style={draw=figblue, rounded corners=3pt, fill=figblue!10, align=center, inner sep=4pt},
      open/.style={state, fill=figblue!25},
      act/.style={font=\footnotesize\ttfamily, align=center, text=figamber},
      arr/.style={-{Stealth[length=2.2mm]}, figslate},
      lbl/.style={font=\footnotesize\itshape, figslate},
    ]
    \node[open] (decl) at (0,0)
      {\textbf{declaration}\\\footnotesize name, typed signature,\\\footnotesize paradigm, capabilities,\\\footnotesize $\alpha$, resources};
    \node[open, right=1.05cm of decl] (carry)
      {\textbf{open module}\\\footnotesize \texttt{body} $=\varnothing$, valid,\\\footnotesize serializable,\\\footnotesize algorithm is complete};
    \node[state, right=1.05cm of carry] (closed)
      {\textbf{bound module}\\\footnotesize body $=$ call of\\\footnotesize implementation, closed\\\footnotesize for export / execution};
    \node[state, below=0.75cm of carry] (inspect)
      {\textbf{gap report}\\\footnotesize open slots\\ \footnotesize (paradigm, path,\\\footnotesize signature, attributes)};
    \node[act, below=0.9cm of decl] (a1) {abstract\_*\\(declare)};
    \node[act, below=0.9cm of closed] (a2) {bind\\(link)};

    \draw[arr] (decl) -- node[lbl, above, midway] {carried} (carry);
    \draw[arr] (carry) -- node[lbl, above, midway] {attach} (closed);
    \draw[arr] (carry) -- (inspect);
    \draw[arr] (inspect.west) to[bend left=22] (decl.south east);
    \node[lbl] at (1.6, -3.15) {still open};
    \draw[arr] (closed.north) to[bend right=35] node[lbl, above, midway] {re-bind another realization} (carry.north);
    \draw[arr, dashed] (a1) -- (decl);
    \draw[arr, dashed] (a2) -- (closed);
  \end{tikzpicture}
  \caption{The oracle lifecycle.  Declarations create open modules that are
  carried inside otherwise-complete programs. Gap analysis reports what
  remains open. The binding pass attaches implementations (partially and
  repeatedly, with compatibility checks and QRAM resource hoisting) until
  the program is closed for export or execution.}
  \label{fig:oracle-lifecycle}
\end{figure}
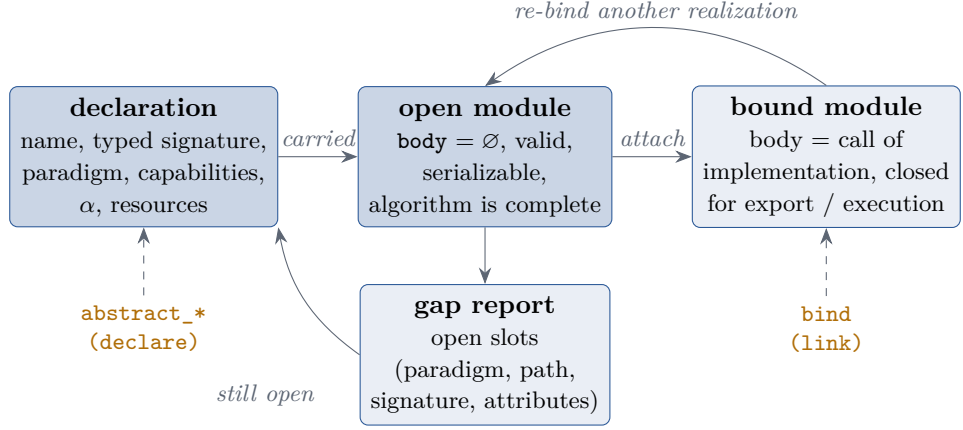

Generation can compose these declarations into a larger program.
Gap analysis reports the reachable unresolved slots and their call paths.
The program can be validated and serialized while open, but execution and
backend export require closure of the entry's reachable dependencies.
An unresolved declaration is therefore neither a missing symbol nor an
operation that silently acts as the identity.
Figure~\ref{fig:oracle-lifecycle} summarizes its lifecycle.

Binding attaches an implementation through a wrapper at the original
slot, retaining the algorithm's existing calls. The operation is
non-mutating. A rejected binding leaves its input program unchanged.
Successful binding may be partial, and the supplied implementation may
introduce additional open dependencies.
The interface checks compare ordered register kinds and widths, declared
roles, and the controlled/adjoint capabilities demanded by callers.
Fixed-point format attributes and a declared block-encoding scale must
also agree. Register names inside the implementation may differ, but its
public argument layout cannot change.

The distinction between structural compatibility and mathematical
correctness is important. A binding report does not prove that a circuit
encodes the intended matrix, that a supplied spectral bound is tight, or
that arbitrary workspace is clean. Explicit contradictory zero-input or
clean-workspace promises are rejected. Missing promises on legacy
operations are not converted into proofs. Such properties require
separate numerical witnesses or mathematical arguments.

\paragraph{Resource capture across linking steps.}
An oracle implementation may require a QRAM that was absent from the
abstract slot. A Binding maps its formal resource to an entry-level
logical name. The linker propagates captured resource arguments along
the call graph rather than embedding the memory contents in RIR.
Capture provenance survives serialization, so later bindings can reuse
the same memory without duplicating intermediate resource parameters.
Independent captures receive a canonical logical-name order.
This matters both for reproducible artifacts and for attributing queries
to the correct logical data source.

\begin{lstlisting}[style=pyqpython, caption={Restore one open algorithm and choose a QRAM realization. The fixture is defined in Section~\ref{sec:oracle-study}. Changing the selected binding leaves the restored algorithm artifact unchanged.}, label={lst:binding}]
from oracq import bind_with_report, dumps, loads
from oracq.applications.oracle_study import oracle_study

opened, variants = oracle_study(width=2, repetitions=3)
restored = loads(dumps(opened))
implementation, memory = variants["qram_table"]
linked = bind_with_report(
    restored, {"AngleWord": implementation}
)
report = linked.report
print(report.ok)                 # True
print(report.bindings)           # (('AngleWord', 'qram_xor_2_2'),)
print(report.resource_mappings)  # (('AngleWord', (('table', 'angle_words'),)),)
closed = linked.require()
\end{lstlisting}

The binding result contains the already-linked program and a structured
report. The report lists source and result digests, chosen
implementations, resource mappings, remaining slots, and any failed
checks with their call paths.
Memory contents remain separate runtime inputs.
Widths, scales, and assembly-time angles are not arbitrary late-bound
symbols. Changing them requires regenerating the algorithm or
constructing an explicit adapter.

\subsection{Role views, protocols, and algorithm contracts}
\label{sec:protocols}

Three objects connect an application to a generated program
(Table~\ref{tab:interfaces}). A Python structural protocol says how an
object supplies an access role. A concrete role view interprets an
operation's registers and associated parameters. An algorithm contract
checks whether those resolved views satisfy a particular generator's
requirements. None of these objects substitutes for the stored RIR.

For example, a matrix input can supply the \texttt{block\_encoding()}
method without inheriting from a framework base class. The adapter below
resolves that method and checks that it returns a \texttt{BlockEncoding}
view. The identity construction keeps the example small. A scientific
input would return its own encoding and normalization.

\begin{lstlisting}[style=pyqpython, caption={A minimal structural-protocol provider. The returned role view is resolved and checked at the Python generation boundary.}, label={lst:be-protocol}]
from oracq import identity
from oracq.algorithms.input_model.interfaces import as_block_encoding

class MatrixInput:
    def block_encoding(self):
        return identity(1)

view = as_block_encoding(MatrixInput())
print(view.alpha)
# 1.0
\end{lstlisting}

\begin{table}[t]
\centering
\caption{Responsibilities at the generation boundary. All Python-side
resolution happens before backend execution.}
\label{tab:interfaces}
\small
\begin{tabular}{@{}p{3cm}p{9.9cm}@{}}
\toprule
Object & Responsibility \\
\midrule
Structural protocol & Supplies an access method, such as
\texttt{block\_encoding()} or \texttt{state\_preparation()}, without
requiring inheritance from a framework base class. \\
Role view & Carries a concrete operation, register interpretation,
normalization or format, and effective call capabilities. \\
AlgorithmContract & Resolves inputs, checks role-specific and cross-input
conditions, and returns the actual checked views with a diagnostic report. \\
RIR declaration & Persists the unresolved quantum implementation after
Python-side generation has finished. \\
\bottomrule
\end{tabular}
\end{table}

One application object can supply multiple roles. A linear-system model
may expose both block-encoding and sparse access, while different solvers
request different views. A concrete view is accepted directly as well.
This reduces coupling between application data structures and solver
implementations without introducing another IR instruction or global
role registry.

Resolution checks that the provider method exists, is callable, accepts
a no-argument invocation when its signature is inspectable, and returns
the required view. A solver generation uses each resolved provider once
and consumes those same checked objects. Independently running a
preflight check and then invoking a solver performs two resolutions.
There is no hidden global cache. Exceptions raised inside provider
methods retain their original cause.

Contracts additionally check widths, domains, capability requirements,
and relations between inputs. For instance, a linear-system solver needs
compatible matrix and right-hand-side target spaces, and a state
preparation suitable for the coherent calls used by that construction.
Spectral or dissipativity promises carry provenance but remain caller
assumptions unless independently justified.
Failures report the input path and expected versus actual conditions,
rather than becoming late backend errors.

Solver objects such as QLSSSolver and QODESolver occupy composable
generator positions. Their output is an operation or state oracle that
can itself be consumed by a larger algorithm.
A success-conditioned state oracle is not automatically a deterministic
clean state preparation. Signal registers and recovery information remain
part of its interface. This explicit boundary is essential when combining
scientific solvers.

\section{Constructing scientific-computing oracles}
\label{sec:primitives}

Scientific algorithms consume linear maps and data-access procedures,
whereas their implementations manipulate register words and quantum
amplitudes. The framework connects these levels through block-encoding
algebra, QRAM resources, and reversible numerical computation.
This section describes the block-encoding layer. The two data-side
mechanisms, QRAM resources and compiled reversible arithmetic, are the
subject of \secref{sec:arithmetic}.
These constructions are library components over RIR, so their calls,
open dependencies, and resource identities remain available to the linker
and estimator.

\subsection{Block-encodings and coherent data access}
\label{sec:be-algebra}
\label{sec:dataloading}

A block-encoding of $A$ is a unitary $U_A$ with designated signal states
such that
\[
 (\langle 0^a|\otimes I)U_A(|0^a\rangle\otimes I)=A/\alpha.
\]
A BlockEncoding view records the target and signal layout and $\alpha$,
while the operation graph determines available controlled and adjoint
calls. A view can wrap an open declaration or an implemented module.
Constructing a view checks its interface. The equation above is a
mathematical obligation, not a consequence of the constructor.

Products, scaling, linear combinations, tensor products, direct sums,
and rectangular embeddings compose these objects.
The construction determines how normalization factors and signal spaces
combine. Consequently replacing a matrix implementation with one having a
different $\alpha$ is not a transparent link-time change after LCU weights
or solver angles have already been generated.
Sparse location and value oracles provide another input route, with
explicit conversion to a block-encoding when a solver requires it.

The same separation applies to state preparation. A preparation specifies
the normalized target state produced on its permitted input and the
workspace convention. Application norms remain separate from normalized
amplitudes. This is particularly important when preparing lifted
initial states or recovering a physical solution from a success branch.

Qubitization, QSVT, and Hamiltonian-simulation kernels consume these
interfaces \cite{low2019qubitization,gilyen2019qsvt,berry2015taylor}.
Approximation kernels are explicit generator arguments, allowing an
algorithm assembly to use a different implementation without changing the
RIR instruction set. Their numerical limitations are discussed in
Section~\ref{sec:limitations}.

\section{Compiled classical arithmetic and coherent data access}
\label{sec:arithmetic}

Scientific access models ultimately supply quantum states with prescribed
amplitudes, and the framework provides two complementary, mechanically
similar ways of producing them.
A QRAM resource \emph{moves} coherent data. The program queries a
declared memory whose classical contents are supplied at run time.
A compiled arithmetic oracle \emph{computes} data. A pure Python
numerical formula is compiled at generation time into a reversible
fixed-point circuit.
Both are ordinary RIR operations with declared registers, XOR-update
semantics, and explicit resource identities, so linking, costing, and
verification treat them uniformly.
This section describes both mechanisms and their composition in the QFVM
case study.

\subsection{QRAM as an explicit resource}
\label{sec:qram}

A QRAM resource is a typed declaration of $w_a$ address bits by $w_d$
data bits, both bounded, and it does not store a Python container
inside the program. Its memory map $M$ is supplied at execution, either
as a word array passed by name to the run entry or as a separate typed
binding file kept beside the program. One program can therefore serve
many data instances. Two instructions address the memory. The Load
instruction
has XOR semantics
\[
 |a\rangle|d\rangle\longmapsto |a\rangle|d\oplus M[a]\rangle,
\]
for arbitrary data words $d$, not only zero targets.
For fixed $M$, the operation is self-inverse and supports coherent
controlled queries. Store is the classical random write
$M[a]\leftarrow d$ discussed below. These semantics implement an access
model \cite{giovannetti2008qram}. They do not specify a physical QRAM
architecture.

A module uses formal memory resources. Call arguments and binding capture
associate them with logical memory identities. Two calls to the same
lookup module with different banks remain distinct in the resource
ledger. Multiple formal names bound to one bank refer to the same
memory.
Changing runtime table contents does not require replacing the algorithm
graph, although it changes the mathematical instance and must be recorded
in its validation data.

\paragraph{Pointer-style access.}
Using the two instructions directly leaves all address bookkeeping to
the caller, so the library offers a generation-stage array view on top
of them. A view fixes a row-major shape over the address space of a
declared resource. Subscripts may be constants, register views, or
slices, and pointers carry constant and register offsets. An address
expression $c+\sum_i k_i r_i \bmod 2^{w_a}$, built from subscripts and
offsets, is materialized into a fresh address register by XOR copies,
shift concatenations, and carry additions. The Load or Store is emitted
against that register. An Adjoint replay of the synthesis restores
the workspace, so the module stays unitary. When the expression is
exactly one full-width register, the instruction addresses it directly
and no arithmetic is emitted. Disjoint dimensions that occupy disjoint
bit segments concatenate without an adder. Only genuinely overlapping
sums, such as a quantum base plus a quantum offset, synthesize one.
Nothing but register arithmetic, Load, and Store ever reaches the RIR,
and the gates of the synthesized arithmetic are counted in logical
estimation like any other gates. Listing~\ref{lst:qmem} shows the
interface.

\begin{lstlisting}[style=pyqpython, caption={A two-dimensional QRAM view addressed by two quantum indices. The subscript sugar lowers to register arithmetic plus one Load. The memory contents are bound only at execution, and the ledger counts the query against the bank separately from gates.}, label={lst:qmem}]
from oracq import Builder, QRAM, QMem, UInt, simulate

b = Builder("field", {"field": UInt(2), "cell": UInt(3), "word": UInt(8)},
            {"state": QRAM(5, 8)})
state = QMem(b, "state", shape=(4, 8))
state[b["field"], b["cell"]].load(b["word"])  # word ^= M[field, cell]
op = b.finish()
final = simulate(op.program(), {"state": list(range(32))},
                 initial={"field": 2, "cell": 5})
print(sorted(final.amplitudes.items()))
# [((2, 5, 21), (1+0j))]
print(sum(op.estimate().qram_queries.values()), op.estimate().gate_total)
# 1 30
\end{lstlisting}

The probe declares a $4\times 8$ state table over a $2^5$-cell resource
and fetches the entry at the superposition-addressable index
$(2,5)$, whose row-major address is $2\cdot 8+5=21$. The thirty gates
are the synthesized addressing and its restoration. Resource estimation
reports queries per bank and writes per bank as first-class counters
alongside gate counts, and text backends lower resource declarations and
runtime writes to dedicated extension lines.

Gate tables and QRAM tables can realize the same XOR interface, and
arithmetic provides a third realization when the table entries follow a
computable rule. Which realization is used is a binding decision, not a
program change. The caller declares the interface, and gate-network,
QRAM, and arithmetic implementations can replace one another under the
compatibility checks of \secref{sec:oracle-lifecycle}.
\secref{sec:oracle-study} quantifies this choice for one angle-word
database, comparing a precomputed gate table, a runtime QRAM bank, and a
reversible modular increment with reported Toffoli and query counts.
An angle-word database also illustrates why interfaces must distinguish
data from amplitudes. Querying a stored word $d_j$ and applying a signal
rotation yields a block proportional to $\cos(s d_j/2)$, not to $d_j$
itself. Angle quantization belongs in the numerical reference for that
construction.

\paragraph{Restricted writes and physical costs.}
Store updates a classical cell $M[a]\leftarrow d$ only when address and
data have definite computational-basis values in the current quantum
state. It does not change the qubits and is not defined for superposed
writes. A module containing this effect, including through a callee,
cannot appear in a controlled or adjoint context.
The logical report records writes separately from gates. Zero gate cost
does not mean zero physical memory cost. Runtime Store is also not
supported by every native execution path.

Fault-tolerant QRAM constructions \cite{dalzell2025qram} and noisy
bucket-brigade studies \cite{wang2026qramnoise} address additional hardware
questions.
This separation lets algorithm researchers inspect the access requirements
before choosing a memory architecture.

\subsection{From numerical formulas to reversible circuits}
\label{sec:mathfunc}

Many scientific access models require values to be computed rather than
looked up. Fixed-point arithmetic supplies reversible addition,
multiplication, division, square roots, comparisons, and selections with
explicit output and status registers.
For a supported numerical function $f$, the compiled operation uses an
out-of-place convention,
\[
 |x,y,s,0\rangle\longmapsto
 |x,y\oplus F(x),s\oplus S(x),0\rangle,
\]
where $F$ is the finite-precision function, $S$ records domain and overflow
status, and the final register denotes cleaned private workspace.
The bitwise XOR convention makes the action meaningful on nonzero output
words as well as on the usual initialized output.

\begin{lstlisting}[style=pyqpython, caption={A physical formula becomes a reversible oracle with explicit fixed-point ports and status. The probe prints the entry module's register layout and the logical cost of the generated network. The example is drawn from the executable math-function documentation.}, label={lst:mathfunc}]
from oracq import FixedFormat, compile_function

def pressure(rho, momentum, energy, gamma=1.4):
    velocity = momentum / rho
    return (gamma - 1) * (energy - 0.5 * momentum * velocity)

compiled = compile_function(
    pressure,
    fmt=FixedFormat(width=12, fraction=6),
    constants={"gamma": 1.4},
)
program = compiled.program()
print([(r.name, r.type.width) for r in program.main.registers])
# [('rho', 12), ('momentum', 12), ('energy', 12), ('out', 12), ('status', 2)]
estimate = compiled.operation.estimate()
print(estimate.qubits, estimate.toffoli, estimate.gate_total)
# 2921 9556 90106
\end{lstlisting}

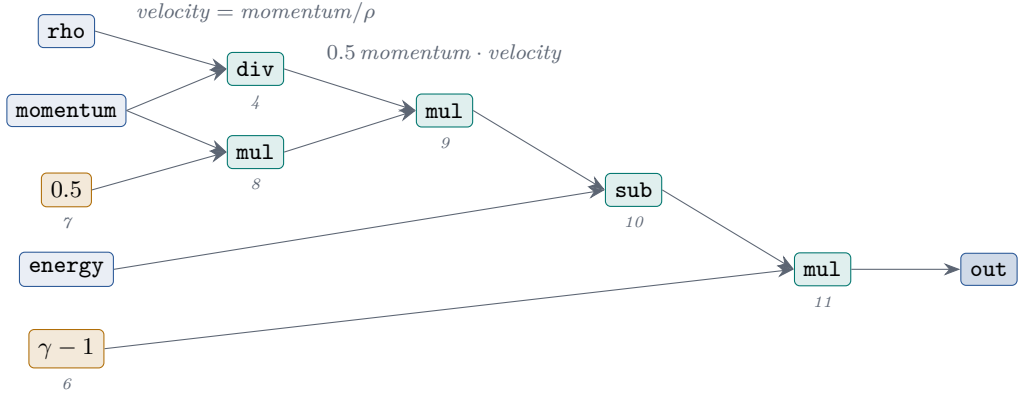
\begin{figure}[t]
  \centering
  \begin{tikzpicture}[
      font=\small,
      var/.style={draw=figblue, rounded corners=2pt, fill=figblue!10, inner sep=3.5pt, font=\footnotesize\ttfamily},
      cst/.style={draw=figamber, rounded corners=2pt, fill=figamber!15, inner sep=3.5pt, font=\footnotesize},
      op/.style={draw=figteal, rounded corners=2pt, fill=figteal!12, inner sep=3.5pt, font=\footnotesize},
      arr/.style={-{Stealth[length=2.2mm]}, figslate},
      idx/.style={font=\tiny\itshape, figslate},
      note/.style={font=\scriptsize\itshape, figslate},
    ]
    \node[var] (rho)      at (0, 2.1)  {rho};
    \node[var] (momentum) at (0, 1.05) {momentum};
    \node[cst, label={[idx]below:{7}}] (half) at (0, 0)     {$0.5$};
    \node[var] (energy)   at (0, -1.05) {energy};
    \node[cst, label={[idx]below:{6}}] (gm1)  at (0, -2.1)  {$\gamma-1$};

    \node[op, label={[idx]below:{4}}] (div) at (2.5, 1.6) {$\mathtt{div}$};
    \node[op, label={[idx]below:{8}}] (mul1) at (2.5, 0.5) {$\mathtt{mul}$};
    \node[note] at (2.5, 2.35) {$\mathit{velocity}=\mathit{momentum}/\rho$};

    \node[op, label={[idx]below:{9}}] (mul2) at (5.0, 1.05) {$\mathtt{mul}$};
    \node[note] at (5.0, 1.8) {$0.5\,\mathit{momentum}\cdot\mathit{velocity}$};

    \node[op, label={[idx]below:{10}}] (sub) at (7.5, 0) {$\mathtt{sub}$};

    \node[op, label={[idx]below:{11}}] (mul3) at (10.0, -1.05) {$\mathtt{mul}$};

    \node[var, fill=figblue!22] (out) at (12.2, -1.05) {out};
    \draw[arr] (mul3) -- (out);

    \draw[arr] (rho.east) -- (div.west);
    \draw[arr] (momentum.east) -- (div.west);
    \draw[arr] (half.east) -- (mul1.west);
    \draw[arr] (momentum.east) -- (mul1.west);
    \draw[arr] (mul1.east) -- (mul2.west);
    \draw[arr] (div.east) -- (mul2.west);
    \draw[arr] (energy.east) -- (sub.west);
    \draw[arr] (mul2.east) -- (sub.west);
    \draw[arr] (gm1.east) -- (mul3.west);
    \draw[arr] (sub.east) -- (mul3.west);
  \end{tikzpicture}
  \caption{The numerical graph compiled from Listing~\ref{lst:mathfunc}
  (the \texttt{math\_ir} attribute of the compiled function, whose node
  numbers are its single-assignment indices). Constant folding has
  evaluated $\gamma-1$ into node~6. Two further constant nodes ($1.4$
  and $1$) remain in the stored graph but are referenced by nothing.
  The graph is
  a tree for this example, although the sequential single-assignment
  form permits sharing. Status does not appear here. It is attached
  during lowering, when each operation node becomes a kernel call whose
  two-bit flag is OR-merged into the status register.}
  \label{fig:arith-tree}
\end{figure}

The probe prints the five registers of the entry module, namely the three
12-bit input words, the out-of-place result word, and the two-bit status
word. It then prints the logical cost of the generated Boolean network,
covering qubits including private workspace, Toffoli gates, and total
gates.
The frontend inspects a supported subset of Python syntax rather than
executing arbitrary Python code. It performs constant folding and bounded
inlining of nonrecursive helpers, then builds a typed static
single-assignment numerical graph, shown for this example in
Fig.~\ref{fig:arith-tree}. The graph has twelve live nodes, namely three
inputs, two generation-time constants, and five arithmetic operations.
The frontend has already folded $\gamma-1=1.4-1$ into node~6.
The graph records arithmetic only. Status is added during lowering.
Every operation node becomes a call of a kernel module that writes a
two-bit flag, and the flags are OR-merged into the status register.

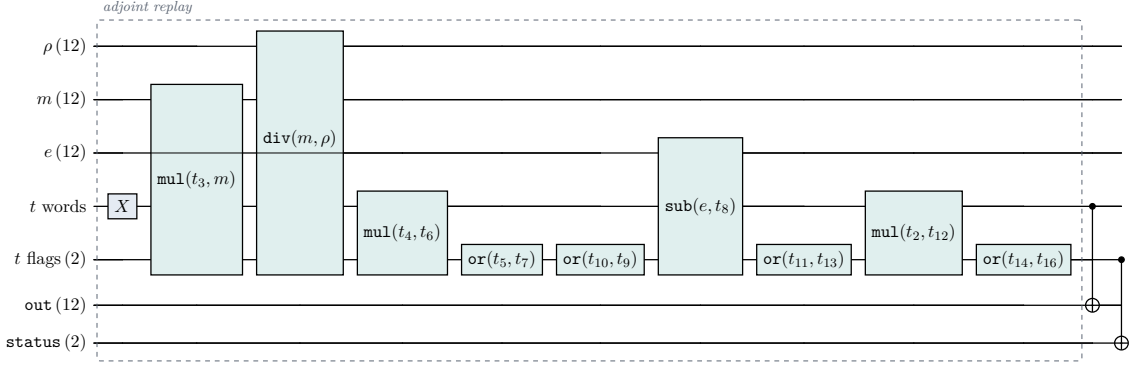
\begin{figure}[t]
  \centering
  \resizebox{\textwidth}{!}{%
  \begin{quantikz}[every matrix/.append style={name=pqsched, column sep=0.8em}, row sep=0.5cm,
      execute at end picture={
        \node[draw=figslate, dashed, rounded corners=2pt, inner sep=3pt,
              fit={(pqsched-col2.west |- pqsched-row1.north)
                   (pqsched-col11.east |- pqsched-row7.south)},
              label={[anchor=south west, font=\footnotesize\itshape, color=figslate]north west:{adjoint replay}}] {};
        \draw (grouppqsched-2-3.west |- pqsched-row3.center) -- (grouppqsched-2-3.east |- pqsched-row3.center)
              (grouppqsched-1-4.west |- pqsched-row3.center) -- (grouppqsched-1-4.east |- pqsched-row3.center);
      }]
    \lstick{$\rho\,(12)$} & \qw & \qw & \gate[wires=5, style={fill=figteal!12}, label style={yshift=0.32cm}]{\mathtt{div}(m,\rho)} & \qw & \qw & \qw & \qw & \qw & \qw & \qw & \qw & \qw \\
    \lstick{$m\,(12)$} & \qw & \gate[wires=4, style={fill=figteal!12}]{\mathtt{mul}(t_3,m)} & \ghost{\mathtt{div}(m,\rho)} & \qw & \qw & \qw & \qw & \qw & \qw & \qw & \qw & \qw \\
    \lstick{$e\,(12)$} & \qw & \ghost{\mathtt{mul}(t_3,m)} & \ghost{\mathtt{div}(m,\rho)} & \qw & \qw & \qw & \gate[wires=3, style={fill=figteal!12}]{\mathtt{sub}(e,t_8)} & \qw & \qw & \qw & \qw & \qw \\
    \lstick{$t$ words} & \gate[style={fill=figblue!12}]{X} & \ghost{\mathtt{mul}(t_3,m)} & \ghost{\mathtt{div}(m,\rho)} & \gate[wires=2, style={fill=figteal!12}]{\mathtt{mul}(t_4,t_6)} & \qw & \qw & \ghost{\mathtt{sub}(e,t_8)} & \qw & \gate[wires=2, style={fill=figteal!12}]{\mathtt{mul}(t_2,t_{12})} & \qw & \ctrl{2} & \qw \\
    \lstick{$t$ flags\,(2)} & \qw & \ghost{\mathtt{mul}(t_3,m)} & \ghost{\mathtt{div}(m,\rho)} & \ghost{\mathtt{mul}(t_4,t_6)} & \gate[style={fill=figteal!12}]{\mathtt{or}(t_5,t_7)} & \gate[style={fill=figteal!12}]{\mathtt{or}(t_{10},t_9)} & \ghost{\mathtt{sub}(e,t_8)} & \gate[style={fill=figteal!12}]{\mathtt{or}(t_{11},t_{13})} & \ghost{\mathtt{mul}(t_2,t_{12})} & \gate[style={fill=figteal!12}]{\mathtt{or}(t_{14},t_{16})} & \qw & \ctrl{2} \\
    \lstick{$\mathtt{out}\,(12)$} & \qw & \qw & \qw & \qw & \qw & \qw & \qw & \qw & \qw & \qw & \targ{} & \qw \\
    \lstick{$\mathtt{status}\,(2)$} & \qw & \qw & \qw & \qw & \qw & \qw & \qw & \qw & \qw & \qw & \qw & \targ{}
  \end{quantikz}%
  }
  \caption{The compute/XOR-update/uncompute schedule of the entry module
  compiled in Listing~\ref{lst:mathfunc}, one column per instruction
  group. The X gate seeds the folded constants ($\gamma-1$ into $t_2$,
  $0.5$ into $t_3$). Here $t_k$ denotes the private local
  \texttt{math\_tmp\_k}, and $(m,e,\rho)$ the input words. A call box
  spans the registers its kernel module reads and writes. Each
  arithmetic call consumes operand words, writes its result word into
  the $t$-word bundle and a two-bit flag into the $t$-flag bundle, and
  a wire crossing a box interior is unaffected. Four logic-or calls
  merge the flags, and two closing XOR copies move $t_{15}$ into
  \texttt{out} and $t_{17}$ into \texttt{status}. The dashed region is
  replayed in reverse by one Adjoint instruction, returning all private
  locals to zero. The input registers $\rho$, $m$, and $e$ are read but
  never modified, consistent with the XOR-update convention.}
  \label{fig:arith-schedule}
\end{figure}

Lowering turns the graph into the entry module's instruction schedule,
sketched in Fig.~\ref{fig:arith-schedule}. Four X gates materialize the
folded constants in private locals. Each arithmetic node is a call into
a shared kernel module with its own two-bit status flag. Four logic-or
calls merge the flags. Two closing XOR copies move the result word into
\texttt{out} and the merged flags into \texttt{status}. One Adjoint
block replays the entire forward section in reverse, restoring every
private local to zero.
All kernel modules obey a zero-in/zero-out workspace contract, and calls
of the same module share one deduplicated RIR definition. At this format
the division, multiplication, subtraction, and status-OR kernels occupy
2805, 1026, 112, and 12 qubits respectively. Because calls are
sequential, their private workspaces share one stretch of physical
qubits, which is why the whole-program estimate in
Listing~\ref{lst:mathfunc} reports 2921 qubits rather than the sum of the
module figures.

Real and complex numerical kernels have documented domains and
approximation choices. Transcendental operations are not exact black
boxes. Unsupported source constructs are rejected.

This pipeline separates three errors that can otherwise be conflated.
These are approximating a mathematical function, representing its
values and constants with finite precision, and implementing the
resulting word-level
operation as a quantum circuit. A compiled pressure function can be exact
relative to its fixed-point specification while still differing noticeably
from real-valued fluid physics. Validation reports these comparisons
separately.

The lowering currently uses Boolean arithmetic networks. QFT-based
comparators and modular arithmetic, including the low-ancilla
constructions of Ref.~\cite{yuan2023comparator}, are relevant
alternative realizations, not the circuits underlying our reported gate
counts.

Finally, this compiler acts on numbers encoded in computational-basis
registers. It does not implement an arbitrary nonlinear transformation
of quantum amplitudes. The nonlinear PDE application instead uses QHAM
to construct a larger linear system, as discussed in
\secref{sec:casestudies}.

\subsection{Compiled flux entries over QRAM-served fields}
\label{sec:arith-qram}

The QFVM case study (\secref{sec:qfvm}) composes the two mechanisms.
Its sparse-entry routine is built from a pure Python function of six
conserved field values, namely the density, momentum, and energy of the
two
cells adjacent to a face, plus a face selector and the four-valued
row/column position of the matrix entry. The same frontend compiles it
into a reversible fixed-point module with two \texttt{Index}-typed
selector ports and dual flux outputs (left and right).
At query time the entry oracle holds source, row, column, and band
addresses in superposition, issues nine coherent loads to the
conserved-variable banks (three fields of a cell and its two neighbours)
and restores them by nine adjoint loads, calls the compiled Roe module
twice for the left and right characteristic contributions, and selects
among three candidate entries according to the band address.
Matrix elements are never pre-stored. The banks hold only raw conserved
variables, and each entry exists only as the XOR-update written by the
compiled arithmetic. When the arithmetic reports a nonzero status, such
as a fixed-point division by zero or an overflow, the element contributes
zero by the documented convention.
\secref{sec:numverify} witnesses this fixed-point physics against an
independent word-level reference, including the measured gap between the
bit-exact circuit semantics and float64 fluid physics.

\section{The QFVM and QHAM scientific workflows}
\label{sec:ode}
\label{sec:casestudies}

The two applications exercise different uses of the framework.
QFVM computes matrix entries from physical field data using reversible
arithmetic and then supplies a quantum linear-system solver.
QHAM first constructs a linear representation of a truncated nonlinear
PDE method, then supplies a quantum evolution solver.
They share access roles, module composition, linking, and evidence
records.

\subsection{Problem objects and solver positions}
\label{sec:ode-inputs}

A linear-system problem specifies a matrix access model, a
right-hand-side preparation and norm, and spectral assumptions such as
$\|A\|\le a$ and $\sigma_{\min}(A)\ge \sigma$ on the intended solve
space. A solver selects the access view its construction needs.
A linear-evolution problem instead supplies a generator $G$, an initial
state preparation and norm, and a time, together with any declared
properties needed by the chosen method.

QODESolver generators provide positions for LCHS
\cite{an2026lchs}, Schr\"odingerization
\cite{jin2024schrodingerization,hu2024schrocircuits}, and CBMD
\cite{wang2026cbmd}. LCHS combines Hermitian-evolution branches.
Schr\"odingerization embeds evolution into an auxiliary coordinate.
CBMD uses a contour decomposition with explicitly chosen nodes and
residues. Each construction exposes its approximation configuration and
the kernel implementing its subproblems. The interface permits
substitution, but the application must still satisfy each method's
conditions and account for its output recovery.

Replacing a solver is a generation-time decision. Replacing an
ABI-compatible oracle body is a link-time decision.
Keeping these operations distinct prevents an algorithm configuration,
a numerical approximation, and a backend realization from being
mistaken for the same kind of substitution.

\subsection{QFVM with field data, reversible fluxes, and a linear solve}
\label{sec:qfvm}

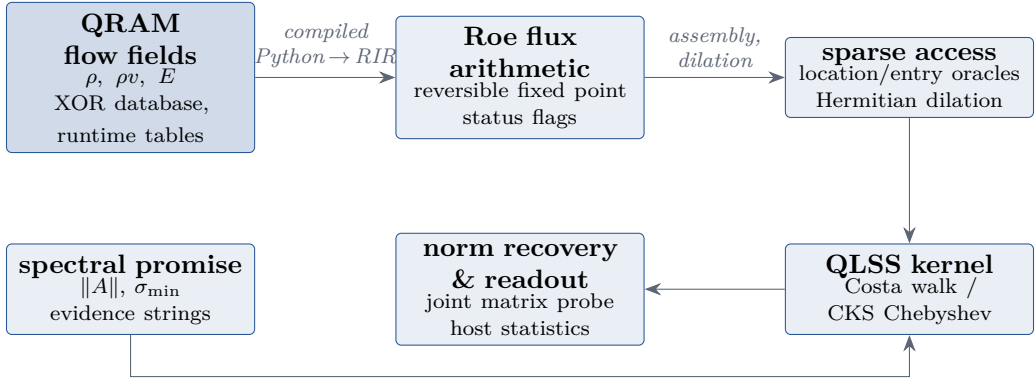
\begin{figure}[t]
  \centering
  \begin{tikzpicture}[
      font=\small,
      stage/.style={draw=figblue, rounded corners=2pt, fill=figblue!10, align=center, inner sep=4pt, text width=3.0cm},
      mech/.style={font=\scriptsize\itshape, text=figslate, align=center},
      arr/.style={-{Stealth[length=2.2mm]}, figslate},
    ]
    \node[stage, fill=figblue!22] (qram) at (0, 0)
      {\textbf{QRAM flow fields}\\[-2pt]\scriptsize $\rho,\ \rho v,\ E$\\[-2pt]\scriptsize XOR database, runtime tables};
    \node[stage] (roe) at (5.15, 0)
      {\textbf{Roe flux arithmetic}\\[-2pt]\scriptsize reversible fixed point\\[-2pt]\scriptsize status flags};
    \node[stage] (sparse) at (10.3, 0)
      {\textbf{sparse access}\\[-2pt]\scriptsize location/entry oracles\\[-2pt]\scriptsize Hermitian dilation};
    \node[stage] (qlss) at (10.3, -2.8)
      {\textbf{QLSS kernel}\\[-2pt]\scriptsize Costa walk /\\[-2pt]\scriptsize CKS Chebyshev};
    \node[stage] (read) at (5.15, -2.8)
      {\textbf{norm recovery \& readout}\\[-2pt]\scriptsize joint matrix probe\\[-2pt]\scriptsize host statistics};
    \node[stage] (prom) at (0, -2.8)
      {\textbf{spectral promise}\\[-2pt]\scriptsize $\|A\|$, $\sigma_{\min}$\\[-2pt]\scriptsize evidence strings};

    \draw[arr] (qram) -- node[above, mech] {compiled\\ Python\,$\rightarrow$\,RIR} (roe);
    \draw[arr] (roe) -- node[above, mech] {assembly,\\ dilation} (sparse);
    \draw[arr] (sparse) -- (qlss);
    \draw[arr] (qlss) -- (read);
    \draw[arr] (prom.south) -- ++(0, -0.55) -| (qlss.south);
  \end{tikzpicture}
  \caption{The QFVM pipeline for the one-dimensional compressible Euler
  equations.  Flow fields are served coherently from a QRAM.  Roe flux
  formulas written as pure Python are compiled to reversible arithmetic.
  The implicit-step operator is exposed as a sparse-access oracle pair.
  The linear solve is an \emph{injected} QLSS kernel operating under an
  explicit spectral promise.  Its fixed-point physics is checked against
  an independent fixed-point semantics in the verification matrix of
  \secref{sec:implementation}.}
  \label{fig:qfvm}
\end{figure}

The quantum finite-volume approach of Ref.~\cite{chen2022qfvm} connects
finite-volume discretization, quantum linear-system solving, and
classical input/output through updateable data access.
Ref.~\cite{zhuang2025navierstokes} develops a related Navier--Stokes
workflow with spectral input/output, explicit circuit synthesis, and
logical-to-physical resource analysis.
Our case study uses these works as algorithmic antecedents.
The presentation follows the data path, showing how conserved field data
reach the arithmetic and how the arithmetic produces matrix entries.
The linear solve is treated as an injected kernel in the position of
\secref{sec:ode-inputs}.
Figure~\ref{fig:qfvm} shows the resulting assembly.

\paragraph{From flux formulas to an implicit linear system.}
In one dimension with periodic cells, the conserved state of cell $j$ is
$U_j=(\rho_j,\rho_j v_j,E_j)$ and the finite-volume semi-discretization
reads
\[
 \frac{\mathrm{d}U_j}{\mathrm{d}t}
   =-\,\frac{F_{j+1/2}-F_{j-1/2}}{\Delta x},
\]
where each numerical flux is the frozen-Roe construction
\[
 F_{j+1/2}=\tfrac{1}{2}\bigl(F(U_j)+F(U_{j+1})\bigr)
   -\tfrac{1}{2}\,\lvert\hat A_{j+1/2}\rvert\,(U_{j+1}-U_j).
\]
The frozen matrix $\lvert\hat A\rvert$ is evaluated at the density-root
weighted averages
\[
 \hat u=\frac{\sqrt{\rho_j}\,u_j+\sqrt{\rho_{j+1}}\,u_{j+1}}
              {\sqrt{\rho_j}+\sqrt{\rho_{j+1}}},
 \qquad
 \hat h=\frac{\sqrt{\rho_j}\,h_j+\sqrt{\rho_{j+1}}\,h_{j+1}}
              {\sqrt{\rho_j}+\sqrt{\rho_{j+1}}},
\]
with sound speed $\hat c^2=(\gamma-1)\bigl(\hat h-\hat u^2/2\bigr)$, and
diagonalized as
$\lvert\hat A\rvert=R\,\operatorname{diag}(a_0,a_1,a_2)\,R^{-1}$, where
the eigenvalues $\lambda\in\{\hat u-\hat c,\ \hat u,\ \hat u+\hat c\}$
are made absolute with the Harten entropy fix
\[
 a(\lambda)=\begin{cases}
   \lvert\lambda\rvert, & \lvert\lambda\rvert\ge\delta,\\[2pt]
   (\lambda^2+\delta^2)/(2\delta), & \lvert\lambda\rvert<\delta.
 \end{cases}
\]
Writing $\Phi^{L}=\tfrac{1}{2}(A_L+\lvert\hat A\rvert)$ and
$\Phi^{R}=\tfrac{1}{2}(A_R-\lvert\hat A\rvert)$ for the left and right
coefficient matrices of a face, with $A_L$ and $A_R$ the Euler flux
Jacobians at its adjacent states, the implicit-step matrix $M$
assembles, per component block,
\[
 M_{j,j-1}=-\frac{\Phi^{L}_{j-1/2}}{\Delta x},\qquad
 M_{j,j}=\frac{\Phi^{L}_{j+1/2}-\Phi^{R}_{j-1/2}}{\Delta x}+\mu,\qquad
 M_{j,j+1}=\frac{\Phi^{R}_{j+1/2}}{\Delta x},
\]
where the optional mass term $\mu$ enters only the component diagonal
(row equals column). The construction satisfies
$MU-\mu U=-r$, the implicit-FVM sign convention relating the operator
to the cell residuals $r$. This identity and the underlying flux
identities are witnessed numerically in \secref{sec:numverify}.
Because $M$ is not symmetric, the solver consumes the Hermitian dilation
$D=\bigl(\begin{smallmatrix}0&M\\ M^{\mathsf T}&0\end{smallmatrix}
\bigr)$ extended by an identity padding block, with output selection
onto the physical coordinate.

\paragraph{Arithmetic over QRAM-served fields.}
The data plane is declared directly in the QRAM primitives of
\secref{sec:qram}. The three conserved fields share one state table, a
QRAM of address width $w_c+2$ and data width equal to the fixed-point
word, viewed with the row-major shape $(3,2^{w_c})$, so that an address
is $\mathrm{field}\cdot 2^{w_c}+\mathrm{cell}$. One entry query holds
the source cell, the component pair, and the band in superposition. For
each neighbour offset in $\{-1,0,+1\}$ it forms the periodic neighbour
address by modular addition on a scratch copy of the cell register, then
fetches the three field words through the two-dimensional pointer
\texttt{state[field, addr].load(word)}. Nine coherent loads, and their
nine adjoint restorations, feed two invocations of the compiled
frozen-Roe face module of \secref{sec:arith-qram}, and the band address
selects the west, center, or east contribution. The geometry table is
the same kind of two-dimensional $(16,2^{w})$ view addressed by
(slot, column), and the residual state is prepared by a QRAM-backed
state-preparation tree described below.
Matrix elements are never pre-stored. The banks hold only raw conserved
variables and packed geometry, and each entry exists only as the
XOR-update written by the compiled arithmetic. When the arithmetic
reports a nonzero status, such as a fixed-point division by zero or an
overflow, the element contributes zero by the documented convention.

\begin{lstlisting}[style=pyqpython, caption={The QFVM physical layer with its data plane declared directly as QRAM resources, using the pointer-style primitives of \secref{sec:qram}. The four-cell fixture executes on the reference simulator. The probe reads one center-block element and prints the query ledger of the generated module.}, label={lst:qfvm}]
from oracq import FixedFormat, simulate
from oracq.applications.qfvm import roe_qfvm_inputs
from oracq.applications.qfvm_qmem import (
    RoeQmemData, qfvm_qmem_physical,
)

inputs = roe_qfvm_inputs(fmt=FixedFormat(5, 2), angle_width=6)
entry = qfvm_qmem_physical(inputs)
data = RoeQmemData([(1, 0, 1), (1, 0.25, 1),
                    (1.25, 0, 1), (1.25, -0.25, 1)],
                   fmt=inputs.fmt, angle_width=inputs.angle_width)
final = simulate(entry.program(), {"state": data.state_bank},
                 initial={"source": 0, "row": 1, "col": 1, "band": 1},
                 max_steps=10_000_000)
(key, _), = final.amplitudes.items()
print(inputs.fmt.decode(key[4]), key[5])  # 1.5 0
print(sum(entry.estimate().qram_queries.values()))  # 18
\end{lstlisting}

The decoded element $1.5$ is the momentum--momentum center entry of
cell zero, that is, the flux couplings of its two interfaces together
with the unit mass term. The zero status records that no arithmetic node
overflowed or divided by zero. The ledger reports eighteen queries
against the state bank, namely the nine coherent loads and their adjoint
restorations.
The same construction exists as a slot-binding path in which the field
banks are open database slots closed by binding to table-realization
modules. That path carries the QLSS problem assembly used in the
profiles of \secref{sec:qfvm-cost}. The pointer-style data plane is its
verified rewrite. Superposition probes of the geometry and state tables,
full-column position permutations, residual amplitudes, and this
physical layer agree amplitude by amplitude with the slot-binding
construction in the repository test suite.

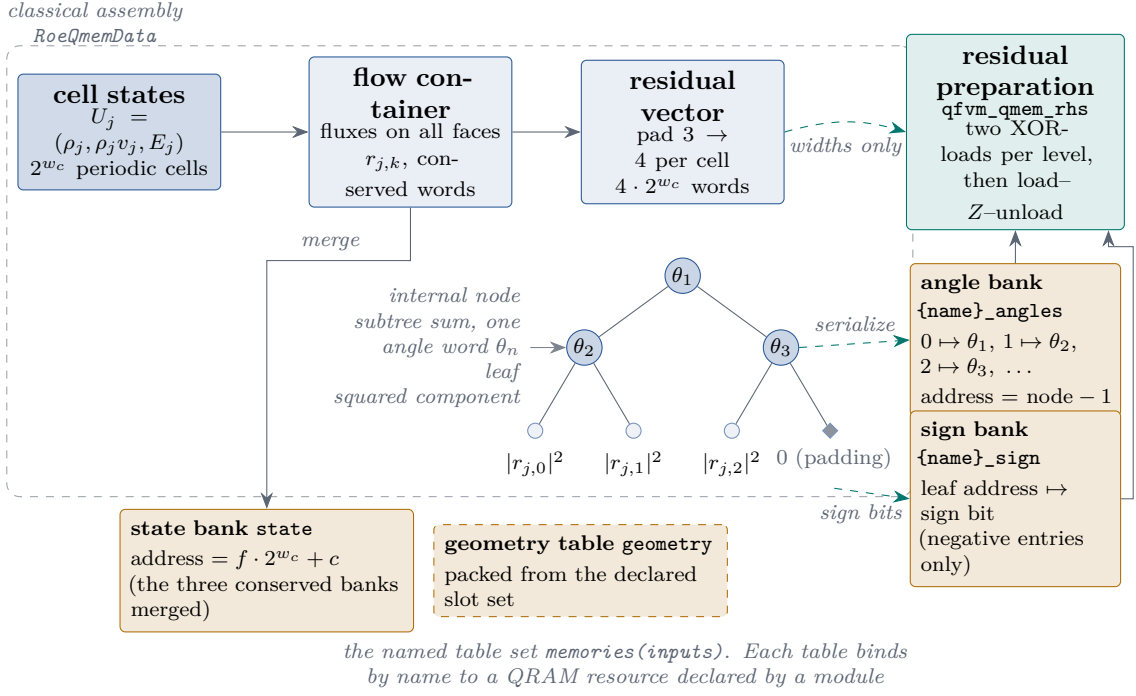
\begin{figure}[t]
  \centering
  \begin{tikzpicture}[
      font=\small,
      stage/.style={draw=figblue, rounded corners=2pt, fill=figblue!10, align=center, inner sep=4pt, text width=2.4cm},
      bank/.style={draw=figamber, rounded corners=2pt, fill=figamber!15, align=left, inner sep=4pt, font=\scriptsize, text width=2.5cm},
      inode/.style={draw=figblue, fill=figblue!25, circle, inner sep=1pt, font=\scriptsize},
      leafn/.style={draw=figblue!65, fill=figblue!8, circle, inner sep=2pt},
      mech/.style={font=\scriptsize\itshape, text=figslate, align=center},
      arr/.style={-{Stealth[length=2.2mm]}, figslate},
      tarr/.style={arr, dashed, draw=figteal},
    ]
    \node[stage, fill=figblue!22] (cells) at (0, 0)
      {\textbf{cell states}\\[-2pt]\scriptsize $U_j=(\rho_j,\rho_j v_j,E_j)$\\[-2pt]\scriptsize $2^{w_c}$ periodic cells};
    \node[stage] (flow) at (3.85, 0)
      {\textbf{flow container}\\[-2pt]\scriptsize fluxes on all faces\\[-2pt]\scriptsize $r_{j,k}$, conserved words};
    \node[stage] (qvec) at (7.45, 0)
      {\textbf{residual vector}\\[-2pt]\scriptsize pad $3\to4$ per cell\\[-2pt]\scriptsize $4\cdot2^{w_c}$ words};
    \draw[arr] (cells) -- (flow);
    \draw[arr] (flow) -- (qvec);

    \node[inode] (root) at (7.45, -1.9) {$\theta_1$};
    \node[inode] (na) at (6.15, -2.85) {$\theta_2$};
    \node[inode] (nb) at (8.75, -2.85) {$\theta_3$};
    \node[leafn] (l1) at (5.5, -3.95) {};
    \node[leafn] (l2) at (6.8, -3.95) {};
    \node[leafn] (l3) at (8.1, -3.95) {};
    \node[leafn, regular polygon, regular polygon sides=4, rotate=45, fill=figslate!70, inner sep=1.4pt] (l4) at (9.4, -3.95) {};
    \draw[figslate] (root) -- (na) (root) -- (nb);
    \draw[figslate] (na) -- (l1) (na) -- (l2) (nb) -- (l3) (nb) -- (l4);
    \node[font=\scriptsize, below=1pt of l1] (lab1) {$\lvert r_{j,0}\rvert^2$};
    \node[font=\scriptsize, below=1pt of l2] (lab2) {$\lvert r_{j,1}\rvert^2$};
    \node[font=\scriptsize, below=1pt of l3] (lab3) {$\lvert r_{j,2}\rvert^2$};
    \node[font=\scriptsize, text=figslate, below=1pt of l4] (lab4) {$0$ (padding)};
    \node[mech, text width=2.7cm, align=flush right] (treeinfo) at (3.95, -2.85)
      {internal node\\ subtree sum, one angle word $\theta_n$\\ leaf\\ squared component};
    \draw[arr, figslate!80] (treeinfo.east) -- (na.west);

    \begin{scope}[on background layer]
      \node[draw=figslate!60, dashed, rounded corners=5pt, inner sep=4pt,
            fit=(cells)(flow)(qvec)(root)(lab1)(lab4)] (roebox) {};
    \end{scope}
    \node[mech, anchor=south west, inner sep=0pt] at (roebox.north west)
      {classical assembly\\ \texttt{RoeQmemData}};

    \node[stage, draw=figteal, fill=figteal!12, text width=2.6cm] (prep) at (11.85, 0)
      {\textbf{residual preparation}\\[-2pt]\scriptsize \texttt{qfvm\_qmem\_rhs}\\[-2pt]\scriptsize two XOR-loads per level,\\[-2pt]\scriptsize then load--$Z$--unload};
    \node[bank] (angles) at (11.85, -2.75)
      {\textbf{angle bank}\\[1pt]\texttt{\{name\}\_angles}\\[2pt]
       $0\mapsto\theta_1$, $1\mapsto\theta_2$,\\ $2\mapsto\theta_3,\ \ldots$\\[2pt]
       address $=$ node${}-1$};
    \node[bank] (signs) at (11.85, -4.85)
      {\textbf{sign bank}\\[1pt]\texttt{\{name\}\_sign}\\[2pt]
       leaf address $\mapsto$ sign bit\\ (negative entries only)};

    \draw[tarr] (qvec.east) to[bend left=16] node[below=2pt, mech] {widths only} (prep.west);
    \draw[arr] (angles.north) -- (prep.south);
    \path (prep.south east) -- (prep.south) coordinate[pos=0.15] (pent);
    \coordinate (jog) at ($(prep.south)!0.5!(angles.north)$);
    \coordinate (lane) at ($(signs.east)+(0.16,0)$);
    \draw[arr] (signs.east) -- (lane) -- (jog -| lane) -- (jog -| pent) -- (pent);

    \draw[tarr] (nb.east) -- node[above=1pt, mech] {serialize} (angles.west);
    \draw[tarr] ($(lab4.south)+(0,-0.12)$) -- node[below=1pt, mech, pos=0.35] {sign bits} (signs.west);

    \node[bank, text width=3.6cm] (statebank) at (1.95, -5.8)
      {\textbf{state bank} \texttt{state}\\[2pt]
       address $=f\cdot 2^{w_c}+c$\\ (the three conserved banks merged)};
    \node[bank, dashed, text width=3.6cm] (geom) at (6.1, -5.8)
      {\textbf{geometry table} \texttt{geometry}\\[2pt]
       packed from the declared slot set};
    \draw[arr] (flow.south) -- (3.85,-1.7) -| node[above, mech, pos=0.28] {merge} (statebank.north);

    \node[mech, text width=12cm] at (6.7, -7.05)
      {the named table set \texttt{memories(inputs)}. Each table binds by name to a QRAM resource declared by a module};
  \end{tikzpicture}
  \caption{The classical assembly behind the QFVM data plane.  The
  dashed container is \texttt{RoeQmemData}. It builds the flow-field
  container from the cell states, pads the quantized residuals to four
  components per cell (one cell's leaves are shown and the padding leaf
  is zero), and raises them to a residual vector whose squared-norm tree
  carries one quantized angle word per internal node.  The tree exists
  on both sides of the boundary. Classically it is an array in which a
  field update rewrites one leaf and its path to the root. Quantumly it
  becomes the QRAM banks serialized from it, namely the angle bank at
  address $\mathrm{node}-1$ and the sparse sign bank at leaf addresses.
  The
  preparation \texttt{qfvm\_qmem\_rhs} is generated from the vector's
  widths and queries the two tree banks at runtime (two XOR-loads per
  level, then load--$Z$--unload for the sign). The merged state bank
  and the slot-set geometry table complete the named table set
  \texttt{memories} consumed by the entry arithmetic of
  Listing~\ref{lst:qfvm}.}
  \label{fig:qfvm-data}
\end{figure}

\paragraph{Assembling the classical data plane.}
Every runtime table of the direct path is exported by one classical
object, \texttt{RoeQmemData}, whose construction is summarized in
Figure~\ref{fig:qfvm-data}.  Given the cellwise triples $U_j$, the
shared fixed-point format, the angle width, and the Roe parameters, the
constructor first instantiates the flow-field container. It validates
the grid, computes the frozen-Roe fluxes on every periodic face, forms
the residuals $r_{j,k}=(F_{j-1/2,k}-F_{j+1/2,k})/\Delta x$ round-tripped
through the fixed-point format, and keeps the encoded conserved words
in three classical banks.  The three physical residual components of
each cell are then padded by a zero fourth component, giving
$4\cdot 2^{w_c}$ fixed-point words, one (component, cell) address per
leaf.  These words are finally wrapped in a residual vector object
(\texttt{QVector}) that builds the squared-norm tree over the decoded
squares and quantizes one rotation angle word per internal node.
The tree is the hinge between the classical and quantum data planes,
because it exists on both sides.  Classically it is a persistent array.
A local field change rewrites one leaf and recomputes only the
leaf-to-root path, so serving an updated field is a bank write, not a
regeneration of the preparation network.  Quantumly the tree is not an
array at all but two QRAM banks serialized from it. The angle bank
holds the internal-node angle words at address $\mathrm{node}-1$, and
the sparse sign bank holds sign bits at leaf addresses.  The
preparation generated from the vector's widths queries these banks at
runtime and never reads the classical array, under the same named-table
binding that serves the state table of Listing~\ref{lst:qfvm}.
Exporting the object therefore yields one named table
set, namely \texttt{state} (the merged $(f,c)$ bank), \texttt{geometry}
(packed from the declared slot set), and the two tree banks, while
\texttt{qfvm\_qmem\_rhs} returns the bank-querying preparation.

\paragraph{Right-hand side and readout.}
The right-hand side is the normalized residual state
\[
 \lvert b\rangle\ \propto\ \sum_{j,k} r_{j,k}\,\lvert j,k\rangle ,
\]
prepared by a squared-norm tree over the residual components. Each
internal node stores the rotation angle
$\theta_n=2\arccos\sqrt{S_{\mathrm{L}}(n)/S(n)}$, quantized into an
angle word, where $S(n)$ and $S_{\mathrm{L}}(n)$ are the squared-norm
sums of a subtree and of its left child. One target bit is decided per
tree level by two XOR-loads of the angle bank, replacing leaf-by-leaf
preparation. Signs are applied by phase kickback. The sign bit is
XOR-loaded into a flag, a $Z$ acts on the flag, and the un-query
restores it, so negative components pick up a factor $-1$ with every
workspace returned to zero.
Norm recovery and the required measurements remain separate from
assembling the coherent solve.

The implementation exposes three replaceable boundaries around this
data path. \emph{Data access} associates named field and geometry
resources with runtime tables, so local changes to the field data are
memory updates rather than gate-network regeneration.
\emph{Numerical realization} fixes the arithmetic format and
approximation rules for the Roe formulas. Changing a public format
requires regeneration, while replacing an implementation with the same
interface can use binding.
\emph{Linear solving} selects a solver generator under an explicit
spectral promise. The CKS Chebyshev construction
\cite{childs2017qls} and the Costa walk construction
\cite{costa2022discrete} consume their corresponding access views of the
same problem object.

This separation lets us inspect where a scientific construction spends
resources. The four-cell study uses the same physical input and records
both open and closed costs for two solver configurations.
It also permits independent checks of the Roe routine, sparse address
permutation, matrix entries, padding, and right-hand-side preparation.
Section~\ref{sec:numverify} reports these numerical witnesses.
Section~\ref{sec:qfvm-cost} reports the assembled logical costs.

\subsection{QHAM, from nonlinear PDE to composable linear input}
\label{sec:qham-case}

The QHAM construction of Ref.~\cite{xue2025qham} combines
homotopy analysis \cite{liao2003beyond,liao2012homotopy} with
quantum-compatible linearization. For the supported polynomial
evolution equations, spatial discretization gives
\[
 \dot u=f+Lu+\sum_{r\ge2}B_r u^{\otimes r}.
\]
A finite homotopy expansion introduces corrections $U_0,\ldots,U_m$,
whose sum approximates the physical solution. Quantum-compatible
linearization organizes the required products of these corrections into
a larger state $Y$ satisfying a linear equation
\[
 \dot Y=G_{m,\eta}Y,\qquad
 u_{\mathrm{HAM}}=\sum_{j=0}^{m}U_j .
\]
Here $m$ and the homotopy parameter $\eta$ are algorithm choices
of the truncated system.

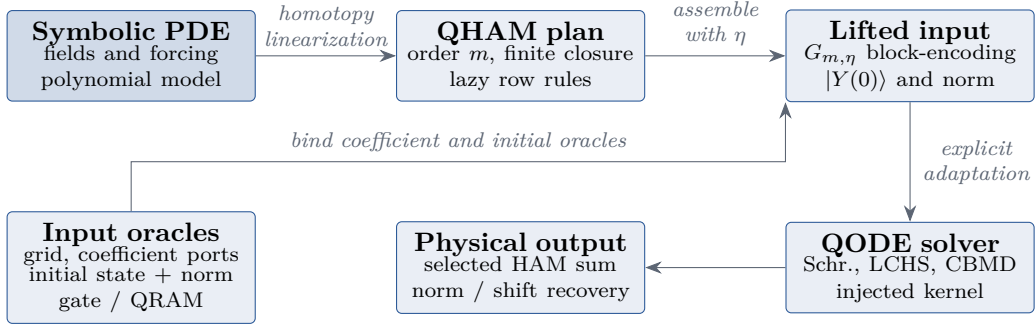
\begin{figure}[t]
  \centering
  \begin{tikzpicture}[
      font=\small,
      stage/.style={draw=figblue, rounded corners=2pt, fill=figblue!10, align=center, inner sep=4pt, text width=3.0cm},
      mech/.style={font=\scriptsize\itshape, text=figslate, align=center},
      arr/.style={-{Stealth[length=2.2mm]}, figslate},
    ]
    \node[stage, fill=figblue!22] (pde) at (0, 0)
      {\textbf{Symbolic PDE}\\[-2pt]\scriptsize fields and forcing\\[-2pt]\scriptsize polynomial model};
    \node[stage] (plan) at (5.15, 0)
      {\textbf{QHAM plan}\\[-2pt]\scriptsize order $m$, finite closure\\[-2pt]\scriptsize lazy row rules};
    \node[stage] (lifted) at (10.3, 0)
      {\textbf{Lifted input}\\[-2pt]\scriptsize $G_{m,\eta}$ block-encoding\\[-2pt]\scriptsize $|Y(0)\rangle$ and norm};
    \node[stage] (solver) at (10.3, -2.8)
      {\textbf{QODE solver}\\[-2pt]\scriptsize Schr., LCHS, CBMD\\[-2pt]\scriptsize injected kernel};
    \node[stage] (physical) at (5.15, -2.8)
      {\textbf{Physical output}\\[-2pt]\scriptsize selected HAM sum\\[-2pt]\scriptsize norm / shift recovery};
    \node[stage] (bindings) at (0, -2.8)
      {\textbf{Input oracles}\\[-2pt]\scriptsize grid, coefficient ports\\[-2pt]\scriptsize initial state + norm\\[-2pt]\scriptsize gate / QRAM};

    \draw[arr] (pde) -- node[above, mech] {homotopy\\linearization} (plan);
    \draw[arr] (plan) -- node[above, mech] {assemble\\with $\eta$} (lifted);
    \draw[arr] (bindings.north) -- (0, -1.4)
      -| node[pos=0.25, above, mech] {bind coefficient and initial oracles} (lifted.south west);
    \draw[arr] (lifted) -- node[midway, right, mech, text width=1.5cm] {explicit\\adaptation} (solver);
    \draw[arr] (solver) -- (physical);
  \end{tikzpicture}
  \caption{The QHAM workflow. The lazy linearization plan and
  discretization/data oracles meet at the lifted QODE input. Coefficient
  realizations and the evolution solver are replaceable. Any required
  dissipative shift and its recovery factor remain explicit. Output
  selection and norm recovery refer to the truncated HAM sum.}
  \label{fig:qham-pipeline}
\end{figure}

The implementation is a composition hierarchy in which each level is an
ordinary program object and each boundary is replaceable independently.
At the \emph{symbolic level}, the PDE is a polynomial expression over
named fields and forcing terms.
At the \emph{linearization level}, a QHAMPlan fixes the truncation
order $m$ and the homotopy parameter $\eta$ and describes the finite
closure and its row rules lazily. No tensor blocks and no full generator
matrix are ever materialized. This mathematical plan is an object
distinct from the RIR that stores the generated quantum operations.
At the \emph{discretization level}, bindings turn a grid and the
coefficient structure into ports. These ports are block-encodings for
the linear part,
the multilinear couplings, and the forcing, plus a repeatable
initial-state preparation. Structured finite-difference bindings use
shifts and local contractions. Small explicit matrix constructions
provide references. Named coefficient databases permit gate-table or
QRAM realizations of the same ports (\secref{sec:qram}).
At the \emph{lifted input level}, the generator encoding of $G_{m,\eta}$
and the lifted initial state are assembled from those ports, and at the
\emph{solver level} any QODE protocol of \secref{sec:ode-inputs}
consumes the assembled model, with output selection returning the
physical HAM sum.
Figure~\ref{fig:qham-pipeline} summarizes how these inputs meet at the
lifted linear model before an evolution solver is selected.
Listing~\ref{lst:qham} realizes the pipeline up to the assembled input,
and Listing~\ref{lst:qham-qode} continues it through the solver level.

\noindent\begin{minipage}{\linewidth}
\begin{lstlisting}[style=pyqpython, caption={The core QHAM construction for a forced Burgers fixture. Imports are given in the complete executable tutorial. Listing~\ref{lst:qham-qode} continues from the assembled model. The PDE plan, discretization, and quantum input model are separate objects.}, label={lst:qham}]
u = Field("u")
pde = PolynomialPDE.from_equations({
    "u": 0.1*u.d("x", 2) - u*u.d("x") + Known("forcing")
})
plan = QHAMPlan(pde, order=2)
grid = Grid(("x",), (4,), (1.0,), boundary="periodic")
space = Discretization(
    pde, grid, {"forcing": [0.05, 0, -0.05, 0]}
)
initial = space.encode_fields({"u": [0.1, 0.2, 0, -0.1]})
ports = structured_fd_bindings(space, initial)
model = qham_input_model(plan, ports, eta=-0.4)
print(plan.max_rank, model.generator.width)
# 3 8
\end{lstlisting}
\end{minipage}

The probe reports the maximum tensor rank of the finite closure (rank
three for this order-two plan) and the qubit width of the assembled
lifted-generator block encoding (the truncated lifted state fits in
eight qubits).
Around the assembled model, two independent composition decisions
remain exposed.
The application can bind the coefficient ports while retaining a common
lifted-system assembly, or keep the entire lifted generator as an open
block-encoding with its mathematical plan recorded separately.
It can then choose a compatible evolution solver without rewriting the
PDE expression. Substitutions at the coefficient level are link-time
decisions, and solver changes are generation-time decisions, the two
operations separated in \secref{sec:ode-inputs}.
The same idea extends to other nonlinear lifts, including
Carleman constructions \cite{liu2021pnas,liu2023carleman}.

Listing~\ref{lst:qham-qode} continues from the assembled model at the
solver level, where QODE protocols appear as ordinary callable objects.
A \texttt{linear\_qode} solver couples one evolution method with a
Hamiltonian-simulation kernel. Handing it to \texttt{model.solve} is the
entire integration step, and the returned oracle already holds the
physical HAM sum after output selection.
Schr\"odingerization accepts the lifted generator as assembled, whereas
the CBMD configuration is applied to an explicitly shifted model whose
recorded factor re-enters through $e^{\mu t}$ in recovery.

\noindent\begin{minipage}{\linewidth}
\begin{lstlisting}[style=pyqpython, caption={QODE solvers injected into the QHAM model of Listing~\ref{lst:qham}. Schr\"odingerization consumes the assembled generator directly, while CBMD runs after the explicit dissipative shift, whose recorded factor is printed first. The loop reports qubit count and Toffoli count of each generated program.}, label={lst:qham-qode}]
from functools import partial

from oracq import estimate_resources
from oracq.algorithms.common.hamiltonian import taylor_hamiltonian
from oracq.algorithms.qode.ode import linear_qode

kernel = partial(taylor_hamiltonian, degree=1)
solution = model.solve(
    linear_qode("schrodingerization", hamiltonian_function=kernel), 0.01
)
shifted = model.dissipative_shift()
print(round(shifted.growth_shift, 2))  # 9.78
cbmd = shifted.solve(
    linear_qode("cbmd", hamiltonian_function=kernel), 0.01
)
for name, state in (("schrodingerization", solution), ("cbmd", cbmd)):
    est = estimate_resources(state.operation.program())
    print(name, est.qubits, est.toffoli)
# schrodingerization 151 68073
# cbmd 151 460621
\end{lstlisting}
\end{minipage}

Both selections return a state oracle for the same physical output
block, so the printed numbers compare generation-time choices rather
than physics. The Schr\"odingerization program uses $68{,}073$ Toffoli
gates against $460{,}621$ for the shifted CBMD program, at $151$
qubits each under the same degree-one Taylor kernel. The printed
$9.78$ is the $\mu$ recorded on the shifted model.
Exchanging one solver for another, or retuning its kernel or
quadrature plan, regenerates the program without touching the PDE
expression, the discretization, or the coefficient bindings.

Initial-state preparation must preserve relative norms between lifted
blocks. If the original vector has norm $r$, its $k$-fold tensor component
has norm $r^k$. Independently normalizing each block would change the
linearized problem. The implementation uses repeatable state-preparation
oracles, not cloning of an unknown input state, and records the overall
initial norm and physical output window.

The choice of evolution method also exposes an important mathematical
boundary. A lifted generator need not be dissipative.
An explicit shift $G_s=G-\mu I$, applied before the solve as in
Listing~\ref{lst:qham-qode}, may supply the condition required by an
LCHS or CBMD configuration, with
$e^{Gt}=e^{\mu t}e^{G_st}$ used in recovery.
The growth factor remains part of the report. Shifting is not a free
improvement of success probability or cost.
Schr\"odingerization instead has auxiliary-grid and recovery-window
conditions.

The QHAM case therefore demonstrates a different benefit from QFVM.
QFVM composes one deep data path in which compiled arithmetic consumes
QRAM-served fields. QHAM instead composes many levels, and a scientific
linearization is represented independently of its coefficient access
and solver realization. Numerical checks of generator blocks,
lifted-state weights, and finite evolution constructions test those
connections directly.

\section{Verification of quantum scientific-computing programs}
\label{sec:verification}
\label{sec:implementation}

Every numerical and resource claim in this paper is a classical statement
about a compiled artifact, so the verification methodology that produces
those statements deserves the same attention as the constructions it
evaluates.
This section first sets out that methodology. It explains why classical
verification
carries the evidence burden for quantum scientific computing
(\secref{sec:verify-why}) and which obstacles shape the comparison
design (\secref{sec:verify-how}).
The remainder of the section then reports the retained evidence,
namely a controlled three-realization experiment, the domain witness
campaign, and the resource analyses.

\subsection{Why classical verification carries the evidence}
\label{sec:verify-why}

A quantum scientific-computing program is not one artifact but a stack of
boundaries, each of which can change numerical meaning silently.
A mathematical construction is lifted into a stored open program. The
open
program is bound to an oracle realization. Formulas are compiled to
reversible fixed-point arithmetic. The closed program is then lowered
to a
backend dialect.
A block-encoding scale applied twice, a register format reinterpreted
between caller and callee, a truncation constant folded differently in a
helper, or a lowering defect in the execution path can each corrupt a
result without producing any error signal.
None of these failures announces itself. The simulation completes and
returns numbers.

Two properties of this framework make verification pressing rather than
optional.
First, oracle realizations are deliberately replaceable, so the same
algorithm routinely exists in several compiled forms. That freedom is
useful only if the forms can be \emph{checkably} equivalent on concrete
instances. Otherwise the framework would multiply unverifiable variants
of every construction.
Second, resource claims are computed from the same lowered programs as
the numerical runs. Gate counts that other researchers might quote are
only as credible as the compilation whose output is counted. An
unverified lowering undermines the circuits and the numbers
simultaneously.

No fault-tolerant device can execute these algorithms at their intended
scale today, so every numerical statement about a compiled construction
rests on classical evidence. Classical verification is therefore not an
auxiliary quality-assurance step but the main instrument through which
such a statement earns credibility. Because program outputs are
success-conditioned quantum states rather than classical arrays, that
evidence must be designed deliberately. One chooses what to compare
(amplitudes, matrix entries, bit patterns, or probability distributions),
against which reference, and with what acceptance criterion.

\subsection{Difficulty, criteria, and execution paths}
\label{sec:verify-how}

\begin{figure}[t]
  \centering
  \begin{tikzpicture}[
      font=\small,
      stage/.style={draw=figblue, rounded corners=2pt, fill=figblue!10, align=center, inner sep=4pt, text width=2.7cm},
      refbox/.style={stage, text width=3.4cm},
      wide/.style={stage, text width=12.6cm},
      cmp/.style={draw=figamber, rounded corners=2pt, fill=figamber!15, align=center, inner sep=4pt, text width=5.4cm},
      mech/.style={font=\scriptsize\itshape, text=figslate, align=center},
      arr/.style={-{Stealth[length=2.2mm]}, figslate},
    ]
    \node[stage, fill=figblue!22] (open) at (0, 0)
      {\textbf{open program}\\[-2pt]\scriptsize stored RIR, unresolved slot};
    \node[stage] (gate) at (3.9, 0)
      {\textbf{gate table}\\[-2pt]\scriptsize truth-table network};
    \node[stage] (qram) at (7.3, 0)
      {\textbf{QRAM table}\\[-2pt]\scriptsize runtime memory};
    \node[stage] (arith) at (10.7, 0)
      {\textbf{arithmetic}\\[-2pt]\scriptsize reversible fixed point};

    \node[wide] (paths) at (5.35, -2.4)
      {\textbf{execution paths}\\[-2pt]\scriptsize reference executor \;\,$\cdot$\;\, PySparQ event adapter\\[-2pt]\scriptsize native PySparQ RIR \;\,$\cdot$\;\, OriginIR-ext / UnifiedQuantum};

    \node[refbox] (iref) at (0, -4.7)
      {\textbf{independent reference} $u_{\theta}$\\[-2pt]\scriptsize word-level semantics\\[-2pt]\scriptsize closed-form amplitudes\\[-2pt]\scriptsize classical integration};
    \node[refbox] (mref) at (8.4, -4.7)
      {\textbf{mathematical reference} $u$\\[-2pt]\scriptsize float64 formulas, exact solutions};

    \node[cmp] (eimpl) at (0, -7.0)
      {$E_{\mathrm{impl}}=\|\widehat u_{\theta}-u_{\theta}\|$\\[-2pt]\scriptsize paths vs.\ reference, $<10^{-9}$ or bit-exact};
    \node[cmp] (emeth) at (8.4, -7.0)
      {$E_{\mathrm{method}}=\|u_{\theta}-u\|$\\[-2pt]\scriptsize reported separately, never merged};

    \draw[arr] (open.east) -- (gate.west);
    \draw[arr] (gate.east) -- (qram.west);
    \draw[arr] (qram.east) -- (arith.west);
    \draw[arr] (open.south) -- (paths.north -| open);
    \draw[arr] (gate.south) -- (paths.north -| gate);
    \draw[arr] (qram.south) -- (paths.north -| qram);
    \draw[arr] (arith.south) -- (paths.north -| arith);
    \draw[arr] (paths.south -| iref.north) -- (iref.north);
    \draw[arr] (iref.south) -- (eimpl.north);
    \draw[arr] (mref.south) -- (emeth.north);
    \draw[arr] (iref.east) -- node[above, mech] {$u_{\theta}$} (3.4, -4.7)
      -- (3.4, -7.0) -- (emeth.west);
  \end{tikzpicture}
  \caption{Verification architecture. One stored open program is bound to
  interchangeable oracle realizations and executed through four backend
  paths. Implementation error is measured against references constructed
  independently of the tested code. Method error against the mathematical
  reference is reported separately and is not reduced by any amount of
  backend agreement.}
  \label{fig:verification-overview}
\end{figure}
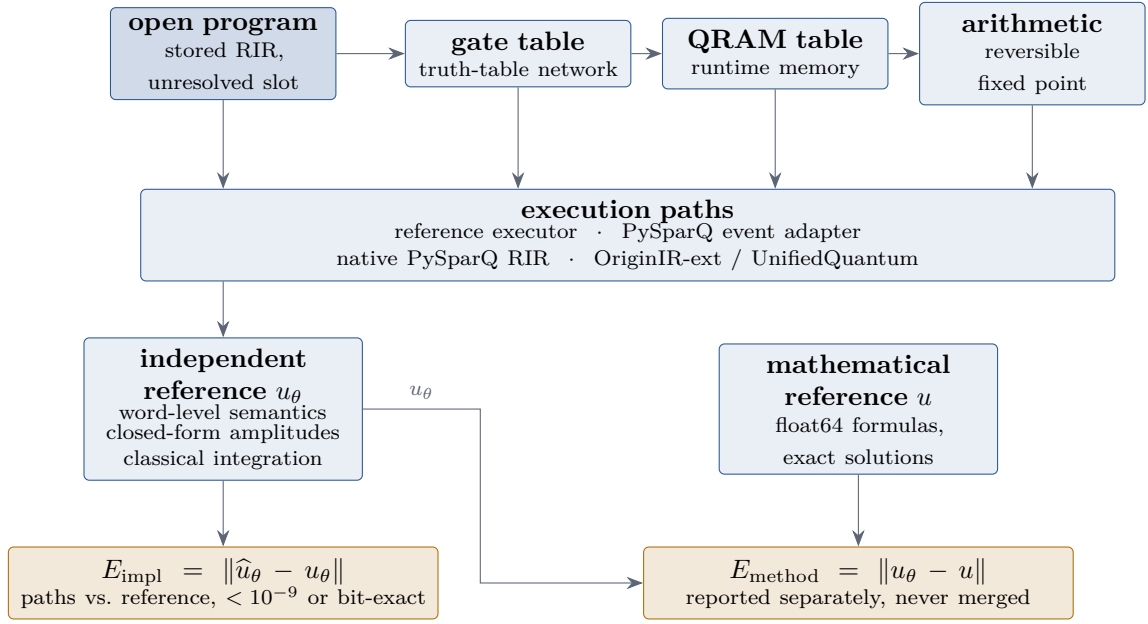

\paragraph{Exponential state spaces.}
Full-amplitude comparison is feasible only for small instances. The
OriginIR-ext state-vector route operates within a 24-qubit budget, and
sparse PySparQ routes extend coverage to structured cases with roughly
30-qubit workspaces. Beyond these sizes, coverage must be budgeted per
case and recorded per case. It is never inferred from group totals.
\figref{fig:verification-overview} summarizes how programs, execution
paths, and reference levels are combined.

\paragraph{Reference independence.}
A comparison against the engine that produced a result is circular. The
campaign therefore uses references constructed independently of the
implementation under test. These are closed-form success and failure
amplitudes,
an independently transcribed word-level fixed-point semantics (with its
own truncation, wraparound, and status conventions), float64
mathematical formulas, and classical integrations of the truncated
method. Different observables test different claims and are reported
separately.
\begin{itemize}
\item maximum amplitude or state-vector deviation between execution
paths and references (typical acceptance $10^{-9}$).
\item bit-level equality of output words and status flags against the
independent fixed-point semantics, for compiled arithmetic.
\item effective-block extraction from full unitaries, with the encoded
matrix $A/\alpha$ and a leakage bound on out-of-block support, for
block-encodings.
\item fidelity and total-variation distance, for preparation and
sampling distributions.
\item distribution-level bind invariance across candidate realizations
of one unresolved slot.
\end{itemize}
These metrics are not merged into a single headline number.

\paragraph{Fixed-point semantics versus floating point.}
A compiled reversible circuit implements exact word-level semantics, not
floating-point arithmetic. The natural reference for implementation
error is therefore an independently rounded word-level oracle, and the
natural criterion is bit-exactness. The float64 evaluation of the same
formula belongs to the method side. The two are separated explicitly
(\secref{sec:numverify} and Appendix~\ref{app:evidence}).

\paragraph{Backends are software too.}
The four execution paths (the Python reference executor, the PySparQ
event adapter, native PySparQ RIR interpretation, and
OriginIR-ext/UnifiedQuantum) provide differential implementation
checks, but they are not four independent numerical engines. The two
PySparQ routes share one simulator. Recorded defects in backend lowering
paths are handled by documented workarounds and regression pins, which
is why cross-path agreement is an observation to be reported with each
run rather than an assumption. Resource accounting is likewise checked
against emitted strict netlists rather than trusted from the estimator
alone.

Evaluation then addresses three questions. It asks whether stored
algorithms
accept compatible realizations, whether the resulting computations agree
with independent references, and how the chosen realizations affect
logical resources.
What the library covers, relative to other frameworks' standard
libraries, is discussed in \secref{sec:related}
(Table~\ref{tab:ode-coverage}). The remainder of this section reports
what is validated, and at what cost.

\subsection{One open algorithm, three realizations}
\label{sec:oracle-study}

A controlled experiment isolates oracle binding from solver choices.
The access contract is
\[
 |a,d\rangle\longmapsto |a,d\oplus k(a)\rangle,\qquad
 k(a)=(a+1)\bmod 2^w .
\]
The algorithm uniformly prepares the address, computes the angle word,
rotates a signal qubit by $2\pi k(a)/2^w$, and uncomputes the word.
It repeats this module three times. One open JSON artifact is restored
and bound separately to a gate table, a QRAM table, and reversible modular
increment with private workspace. The mathematical function and public
interface remain fixed.

For the specified uniform input, the independent success and failure
amplitudes are
\[
 2^{-w/2}\cos\!\left(\frac{3\pi k(a)}{2^w}\right),
 \qquad
 2^{-w/2}\sin\!\left(\frac{3\pi k(a)}{2^w}\right).
\]
All nine configurations ($w=2,3,4$ and three realizations) agree with this
reference to at most $2.5\times10^{-16}$ through each of four execution
paths, namely the Python reference executor, the PySparQ adapter, native
PySparQ
RIR execution, and OriginIR-ext/UnifiedQuantum.

Before binding, the estimator reports three forward and three adjoint
calls of the unresolved slot and unknown implementation workspace.
After binding, Table~\ref{tab:oracle-study} exposes the cost of each
realization without changing the caller.
The report also records Clifford gates, rotations, success probabilities,
program and memory digests, and the binding diagnostics.

\begin{table}[t]
\centering
\caption{Logical costs for three repetitions of one stored algorithm.
Entries are (Toffoli operations, QRAM queries). A QRAM query is an
unpriced access operation, not a zero-cost physical implementation.
Clifford and rotation costs are retained in the artifacts.}
\label{tab:oracle-study}
\begin{tabular}{@{}rrrr@{}}
\toprule
Word width & Gate table & QRAM table & Arithmetic \\
\midrule
2 & (24, 0) & (0, 6) & (0, 0) \\
3 & (216, 0) & (0, 6) & (12, 0) \\
4 & (960, 0) & (0, 6) & (48, 0) \\
\bottomrule
\end{tabular}
\end{table}

\begin{figure}[t]
  \centering
  \includegraphics[width=0.72\linewidth]{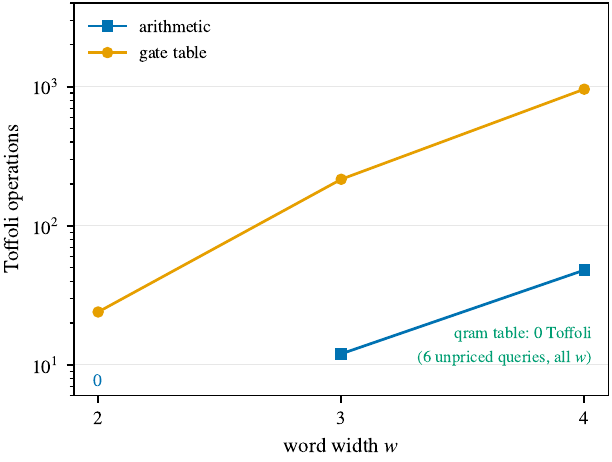}
  \caption{Toffoli cost of the three oracle realizations as the word
  width $w$ grows, for the study of \secref{sec:oracle-study}
  (logarithmic axis, with values from Table~\ref{tab:oracle-study}). The
  arithmetic realization performs zero non-Clifford operations at $w=2$
  and is annotated in place. The QRAM-table realization performs six
unpriced queries instead of Toffoli operations and lies off the axis
at every width.}
  \label{fig:oracle-cost-scaling}
\end{figure}

\figref{fig:oracle-cost-scaling} plots the Toffoli column of
Table~\ref{tab:oracle-study}. The gate-table realization grows
exponentially with the word width because its network enumerates table
entries, while the arithmetic realization grows linearly. The QRAM-table
realization performs six unpriced queries and zero Toffoli operations at
every width, so it does not appear on the logarithmic axis. The zeros are
properties of this lowering, not generic statements about memory or
arithmetic.

The result demonstrates preserved algorithm identity, compatible resource
capture, and implementation-sensitive accounting.
The experiment checks all output branches of one uniform input. Separate
regression tests exercise the arithmetic XOR realization on all two-bit
address/data inputs, including nonzero outputs. The function $k(a)$ was
selected
for its compact arithmetic realization, which arbitrary databases need
not admit.

\subsection{Scientific numerical witnesses}
\label{sec:numverify}

The retained domain campaign consists of 193 cases in six groups
(Table~\ref{tab:verify-matrix}), all passing their stated criteria.
The reports share a source fingerprint and record the tested
configuration, reference, metric, threshold, execution paths, and backend
environment. The nine controlled-study configurations above are additional.

\begin{table}[t]
\centering
\caption{Domain verification supporting the framework and case studies.
A case can contain a basis sweep or multiple checks. Backend coverage is
recorded per case rather than inferred from the group total.}
\label{tab:verify-matrix}
\small
\begin{tabular}{@{}p{3.6cm}rp{8.4cm}@{}}
\toprule
Group & Cases & Independent reference or check \\
\midrule
Oracle algorithms & 30 & Function truth tables and expected oracle actions. \\
Block-encodings & 73 & Encoded matrix blocks, normalization, and finite polynomial constructions. \\
Reversible arithmetic & 20 & Integer permutations and finite-precision numerical references. \\
Compiled numerical functions & 28 & Independent fixed-point semantics, domain/overflow flags, and selected superposition checks. \\
QHAM and QFVM & 15 & Lifted generators and initial states, homotopy identities, Roe entries, sparse locations, and right-hand-side preparation. \\
ODE constructions & 27 & Finite evolution operators, auxiliary-space recovery, and small-instance truncation behavior. \\
\bottomrule
\end{tabular}
\end{table}

\begin{figure}[t]
  \centering
  \includegraphics[width=0.72\linewidth]{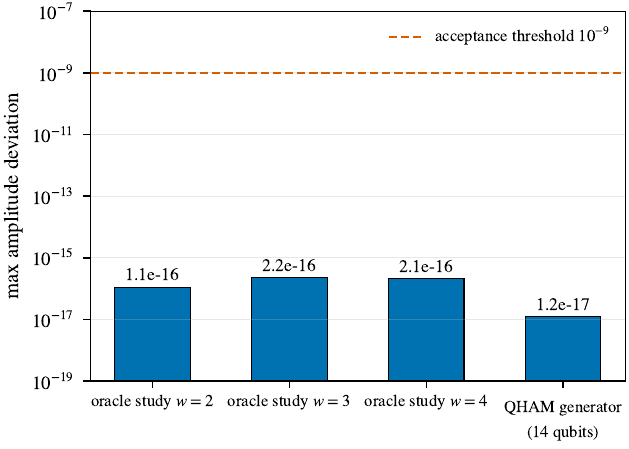}
  \caption{Maximum amplitude deviation across execution paths for the
  controlled oracle study ($w=2,3,4$, worst realization per width) and
  the 14-qubit QHAM generator case, regenerated from the verification
  artifacts. All cases sit orders of magnitude below the $10^{-9}$
  acceptance threshold. The two PySparQ routes share one simulator, so
  these are differential checks rather than four independent engines.}
  \label{fig:multi-path-agreement}
\end{figure}

\paragraph{QHAM.}
For a 28-dimensional lifted-generator fixture, structured and spectral
realizations agree with independently assembled generator matrices to
$1.46\times10^{-17}$ and $7.96\times10^{-18}$, respectively, with no
reported padding leakage. A 14-qubit generator test agrees across the
four execution paths to $8.68\times10^{-18}$. A regenerated run of the
same case reproduces four-path agreement at $1.2\times10^{-17}$, shown
with the realization-study deviations in
\figref{fig:multi-path-agreement}.
In a separate matched-input comparison using $\alpha=4.75$,
structured and spectral realizations have errors below
$1.4\times10^{-17}$, while an eight-bit QRAM angle realization differs by
$6.0\times10^{-5}$, within its stated $9.3\times10^{-3}$ quantization
bound. The nonzero QRAM error is a representation effect, not a failed
backend comparison.

\begin{figure}[t]
  \centering
  \includegraphics[width=0.68\linewidth]{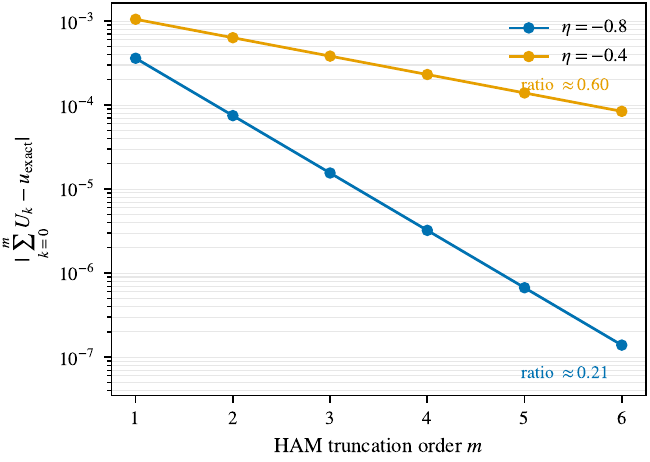}
  \caption{A classical homotopy witness. The plot shows the error of the
  truncated HAM partial
  sum against the exact Riccati solution of the quadratic fixture, for
  truncation orders $m=1,\dots,6$ at $\eta=-0.8$ and $\eta=-0.4$. Both
  series decay geometrically with measured mean ratios matching
  $|1+\eta|$ ($0.21$ and $0.60$). This is a finite mathematical
  construction test, not a quantum simulation of each order.}
  \label{fig:ham-convergence}
\end{figure}

Lifted-state tests check relative tensor-block weights and clean
workspace. A finite first-degree Taylor solve at $t=0.01$ agrees with
its independent truncated reference to $1.12\times10^{-16}$.
A separate classical homotopy witness checks the row rules against the
chain rule and shows decreasing error for orders one through six at
two selected $\eta$ values, plotted in \figref{fig:ham-convergence}.
The measured mean contraction ratios match $|1+\eta|$ at the two
$\eta$ values. These observations support the implemented
lift and finite constructions.

\paragraph{QFVM and compiled physics.}
A Roe-face test at FixedFormat(5,2) matches the independent word-level
reference on all 32 tested branches. A sampled sparse-entry column has
zero implementation error against that reference, but its maximum
difference from real-valued physics is about $0.481$.
The sparse-location permutation is checked on 1,024 branches with no
slot mismatch or dirty workspace. The right-hand-side preparation
agrees with its reference to $1.12\times10^{-16}$.
Classical flux and residual identities are checked independently of the
quantum assembly.

\begin{figure}[t]
  \centering
  \includegraphics[width=0.72\linewidth]{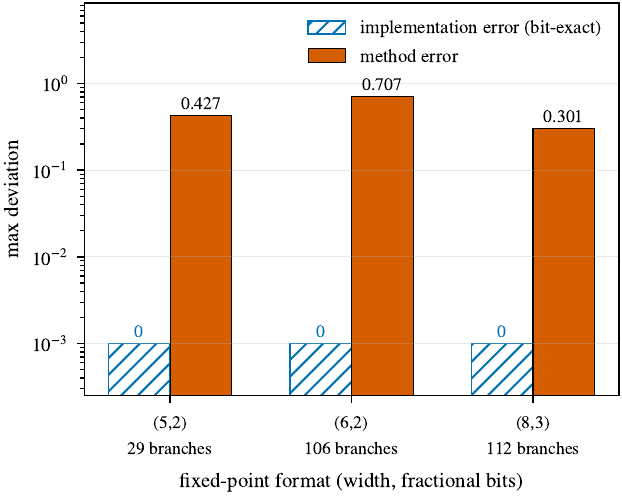}
  \caption{Implementation versus method error for compiled Roe-face
  arithmetic in three fixed-point formats. White hatched bars (drawn at
  the axis floor with true height zero) mark implementation error. The
  circuit output words match the independently rounded semantics
  bit-exactly, so their deviation is $0$. Solid bars give the maximum
  deviation of that fixed-point semantics from the float64 formula on
  the same branches ($0.427$, $0.707$, and $0.301$, with branch counts
  below the
  axis). Backend agreement is exact at every format while the method
  error changes by more than a factor of two.}
  \label{fig:impl-vs-method}
\end{figure}

The compiled-function campaign further separates implementation error
from quantization. For the selected joint Roe-face inputs,
the fixed-point circuit agrees exactly with the independently rounded
semantics in both six-bit/two-fractional-bit and
eight-bit/three-fractional-bit formats.
The corresponding real-valued method errors are about $0.707$ and
$0.301$. \figref{fig:impl-vs-method} shows this separation for the
three formats, including the (5,2) fixture of the domain campaign.
These coarse-format examples illustrate why backend agreement
alone is insufficient evidence of physical accuracy.

\paragraph{Execution coverage.}
Small instances can use all four paths, including dense unitary
extraction where appropriate. Large-workspace arithmetic uses the real
sparse PySparQ routes when the OriginIR simulation budget is exceeded.
For example, the 30-qubit row-preparation case retains both PySparQ
routes, while a 24-qubit lower-precision counterpart exercises four-path
agreement. This is a budgeted coverage distinction, not a replacement
with simulator mocks or an assertion that every case ran on every backend.
Appendix~\ref{app:evidence} defines the reference and error conventions.

\subsection{Resource analysis of open and closed programs}
\label{sec:resources-summary}

Additive logical resources of an open program are represented as
\[
 C(P)=C_{\mathrm{known}}(P)+
       \sum_{s,v}N_{s,v}C_{s,v},
\]
where $s$ is an unresolved slot and $v$ distinguishes controlled,
adjoint, and resource-argument variants.
The report retains the unknown terms rather than replacing them by zero.
A later binding supplies a concrete module graph whose cost can be
evaluated using the same caller structure.

Counts are aggregated per module and call variant, with resource names
substituted on each edge. Repeat multiplies counts without expanding
operations. Rotations are stored by axis, angle, and multiplicity.
Even $2^{40}$ repetitions of a rotation can be represented compactly.
Workspace is instead a peak-live quantity. Open reports give a known
lower bound and list unresolved workspace obligations, not a complete
qubit count.

For closed programs, the estimator follows the actual strict-gate-set
lowering recipes and reports Toffoli, Clifford and exact $T$ operations,
pending rotations, and QRAM reads and writes.
The accounting is checked against emitted strict netlists.
Approximate rotation-synthesis overhead is a declared model rather than
an executed synthesis pass, and the multi-controlled-gate ancilla pool
is reported separately. Detailed conventions are in
Appendix~\ref{sec:resources}.

\subsection{QFVM construction profiles}
\label{sec:qfvm-cost}

The four-cell QFVM fixture uses FixedFormat(6,2), six-bit angle words,
and an explicit spectral promise. Its CKS order-two and Costa one-step
constructions produce the costs in Table~\ref{tab:qfvm-cost}.
Both are generated from the same problem object, then bound to the
same field-data interface and Roe arithmetic.
The report contains open-slot requirements, the closed program and
memory fingerprints, and separate norm-probe artifacts.

\begin{table}[t]
\centering
\caption{Logical profiles of two QFVM constructions. The MCX ancilla
pool is additional to the data/work qubits. These configurations are not
accuracy-matched solver benchmarks or certified CFD time steps.}
\label{tab:qfvm-cost}
\begin{tabular}{@{}lrrrr@{}}
\toprule
Construction & Data/work qubits & MCX pool & Toffoli operations & QRAM queries \\
\midrule
CKS, order 2 & 1,961 & 6 & 58,537,555 & 1,306 \\
Costa, one step & 1,966 & 18 & 649,402,637 & 4,826 \\
\bottomrule
\end{tabular}
\end{table}

Resource attribution remains visible below the solver total.
The CKS construction uses 720 geometry queries, 192 each to density,
momentum, and energy, eight to right-hand-side angles, and two to signs.
The corresponding Costa counts are 2,520, 672 for each conserved-field
bank, 232, and 58. Such differences identify which access realizations
a researcher should examine.

The counts describe this compiler's construction, including known
nonoptimal truth-table and constant-addition lowering.
They do not reproduce the spectral realization or physical-resource
projection of Ref.~\cite{zhuang2025navierstokes}.

\section{Discussion and conclusions}
\label{sec:discussion}
\label{sec:limitations}

The starting point of this work is that in quantum scientific computing
the realization of an access model is itself a research variable. It
changes numerical behavior and resources, and every reported conclusion
is attributable to a realization rather than to an algorithm in the
abstract.
\oracq{} makes that variable a managed decision. Algorithms are
generated against declared access interfaces and stored as open
programs. Checked binding replaces compatible realizations without
regenerating the caller. Construction libraries build those
realizations from typed QRAM resources and compiled reversible
arithmetic, together with block-encoding algebra and its kernels. Each
artifact carries the numerical witnesses and logical cost reports
that make a realization's consequences inspectable.
The evidence follows this structure. QFVM demonstrates raw-data access
and compiled reversible physics. QHAM demonstrates a lazy mathematical
lift with replaceable coefficients and evolution solvers. The
controlled realization experiment establishes that one stored algorithm
can support different implementations and resource accounts, and the
witness campaign ties every numerical claim to a recorded
configuration, reference, and criterion.

The remaining boundaries are concrete.
Binding preserves a fixed public interface and declared assembly
conventions. It can change internal circuitry, private workspace, and
captured resources, but it cannot silently revise a block-encoding
scale, register format, or approximation parameter already used in the
caller, and the concrete-width design keeps parameter sweeps a
Python-generation task.
The tests validate finite instances and explicitly chosen truncations.
The bundled QSP phase synthesizer retains its degree guard of 40, behind
which lies a coefficient-representation limit. The monomial coefficient
pipeline loses reliability around degree 40 independently of the synthesis
route. The synthesizer is a replaceable, contract-checked component, and
with the pyqsp-backed replacement~\cite{chuang2021pyqsp} the retained
$\kappa=8$ inversion fixture now reaches its $10^{-2}$ target at degree
585, with a round-trip residual of $1.0\times 10^{-13}$ against the target
polynomial, an approximation error of $1.8\times 10^{-3}$ against $c/x$ on
$|x| \geq 1/8$ at the phase level, and a direction error of
$5.6\times 10^{-13}$ on the $2\times 2$ block-encoding execution. The
target is constructed and validated in the Chebyshev basis. The
coefficient-level library entry rejects the same request explicitly rather
than degrading silently.
The estimator reports the logical circuit implied by the implemented
lowering. Depth scheduling, routing, distillation factories, error
correction, and physical QRAM loading and latency require additional
assumptions.

The natural next steps are therefore controlled precision studies and
stronger approximation kernels on top of the replaceable phase
synthesizer, implementation comparison as a formal feature with unified,
replayable experiment records, a unified physical readout interface,
improved reversible lowering and reversible-arithmetic cost, staged
interface metadata, and narrow interoperability adapters with existing
algorithm libraries. These developments can be evaluated at the same
interfaces rather than requiring another language core.
In summary, \oracq{} provides a modular scientific-algorithm workflow
that generates against access models, persists the algorithm, selects
compatible realizations, and evaluates their numerical behavior and
logical costs.

\section*{Acknowledgments}
This work has been supported by the National Key Research and Development Program of China (Grant Nos. 2023YFB4502500 and 2024YFB4504100), the National Natural Science Foundation of China (Grant No. 12404564), and the Anhui Province Science and Technology Innovation (Grant Nos. 202423s06050001 and 202423r06050002).

\section*{Code availability}
The \oracq{} framework, its documentation, and the experiment sources of
this paper are released under the MIT license at
\url{https://github.com/IAI-USTC-Quantum/oracq}.
The numerical results reported here were obtained with version 0.1.0
of the framework. The documentation is hosted at
\url{https://iai-ustc-quantum.github.io/oracq}.

\appendix
\section{Reproduction and artifact organization}
\label{app:repro}

The \oracq{} repository separates the paper's experiments from the broader
tutorial and algorithm inventory.
Table~\ref{tab:repro} maps the main results to their executable sources.
Complete walkthroughs of QHAM, Carleman, LCHS, and CBMD are collected in
\texttt{docs/tutorials/scientific-workflows.md}, with links to the source
examples and the relevant mathematical references.
The RIR and backend specifications remain in \texttt{docs/reference/}.
The wider algorithm catalog is documented in \texttt{docs/manual/algorithms/}.
They are not additional claims of this paper.

\begin{table}[h]
\centering
\caption{Executable sources for the retained experiments. Outputs are
generated under \texttt{out/}. Their numerical claims are limited to the
configurations and execution paths recorded in each report.}
\label{tab:repro}
\small
\begin{tabular}{@{}p{4cm}p{8.8cm}@{}}
\toprule
Purpose & Source \\
\midrule
Three oracle realizations & \texttt{examples/research\_workflow.py} \\
QFVM open/closed construction profiles & \texttt{tools/build\_qlss\_comparison.py} \\
QHAM construction and solver selection & \texttt{examples/general\_qham.py} \\
Scientific numerical witnesses & \texttt{tools/run\_verification.py}, with the six groups in Table~\ref{tab:verify-matrix} \\
Binding, resource substitution, and compact repetition checks & \texttt{tests/core/test\_research\_workflow.py} \\
Real-backend interface regression & \texttt{tests/integration/} \\
\bottomrule
\end{tabular}
\end{table}

Generation and core tests require Python but no native quantum backend.
The numerical campaign requires an interpreter with PySparQ and
UnifiedQuantum. The following commands use \texttt{<backend-python>} to
denote that interpreter.
\begin{lstlisting}[style=pyqbase, language=bash, caption={Commands for the paper's experiments and structural regression.}]
python -m unittest discover -s tests/core -v
python examples/research_workflow.py
python examples/general_qham.py
python tools/build_qlss_comparison.py

PYTHONPATH=src <backend-python> examples/research_workflow.py \
  --native --widths 2 3 4 -o out/research-workflow-native
PYTHONPATH=src <backend-python> tools/run_verification.py \
  --group oracles --group blockencoding --group arithmetic \
  --group mathfunc --group qham_qfvm --group ode
PYTHONPATH=src <backend-python> -m unittest \
  discover -s tests/integration -v
\end{lstlisting}

The reported numerical runs used Python 3.11.15, PySparQ source commit
\texttt{4e4c9f16}, \texttt{unified-quantum} 0.1.1,
\texttt{uniqc-cppsimulator} 1.0.1, NumPy 2.4.6, and SciPy 1.17.1.
The full backend identifier and build configuration are recorded in
\texttt{backend-revisions.json}. The native PySparQ RIR interface
requires that source build. Installing a package with the same import
name is not by itself sufficient.

Each domain report includes its source fingerprint, verification-script
fingerprint, backend environment, input parameters, reference metrics,
criteria, and per-case path coverage.
The realization study additionally stores open and closed RIR, runtime
memory snapshots, binding reports, and resource accounts.
Program, memory, and script fingerprints are recorded separately.
Identical RIR with different QRAM contents represents a different
numerical instance.
The QFVM profiles likewise preserve the spectral assumptions and
normalization data needed to interpret their counts.

The supporting core/schema campaign passed 391 tests and the native
integration suite passed 41 tests. These regression counts describe
software checks, not additional scientific accuracy guarantees.
The numerical results in the main text refer specifically to the 193
domain cases and nine realization configurations reported there.

\section{Numerical references and logical-cost conventions}
\label{app:evidence}
\label{sec:resources}

\subsection{Separating reference levels}

For a block-encoding, numerical verification extracts the designated
signal block and compares it with the independently specified
$A/\alpha$. Padding support and workspace are checked where the case
requires them. Matching only a normalized state direction would not test
the same claim.
For a preparation, relative amplitudes and phases, success probability,
and workspace conventions are separate observables.

For an approximate scientific construction, let $u$ denote the
mathematical reference, $u_{\theta}$ the independently evaluated
discretized and truncated method at parameters $\theta$, and
$\widehat u_{\theta}$ the recovered circuit result.
Implementation and method error are distinguished as
\[
 E_{\mathrm{impl}}=\|\widehat u_{\theta}-u_{\theta}\|,\qquad
 E_{\mathrm{method}}=\|u_{\theta}-u\|.
\]
The total error is bounded by their sum in the same norm.
Actual reports specify whether the metric is an amplitude maximum,
matrix-entry maximum, normalized direction error, or relative vector
norm. Different metrics are not combined into a single headline
accuracy number.

For compiled arithmetic, $u_{\theta}$ is replaced by an independently
rounded word-level reference with the selected widths, constants, domain
rules, and overflow behavior.
A valid-domain branch, a flagged branch, and an untested domain are
different outcomes. Coarse-format method errors are retained even when
implementation checks pass.
Likewise, an angle-table realization is compared both with its encoded
integer-angle reference and, where appropriate, with the original
coefficient and its quantization bound.

The homotopy-order witness is a classical test of the finite mathematical
construction, not a quantum simulation of every order in that sweep.
The QFVM cost profiles generate complete bound programs, but their large
solver circuits are not included in a full-amplitude CFD accuracy
claim. Per-case execution paths in the JSON reports preserve this
distinction.

\subsection{Counting conventions}

The logical estimator and strict exporter use the same lowering recipes.
Toffoli operations remain a separate resource. They are not silently
converted into $T$ gates. Clifford atoms and exact $T$ gates are counted
as emitted. Angles on the $\pi/4$ grid use exact Clifford+$T$
decompositions, while other rotations are recorded by axis, angle, and
multiplicity.

For an optional approximate rotation-cost estimate, the report uses the
declared model
\[
 T_{\mathrm{est}} =
 T_{\mathrm{exact}}+
 N_{\mathrm{rot}}\max\!\left(1,
 \left\lceil 3\log_2(1/\epsilon_{\mathrm{rot}})\right\rceil\right),
\]
motivated by logarithmic single-qubit synthesis scaling
\cite{ross2016optimal}.
This is not a compiled synthesis circuit or a global algorithm-error
bound. The reported $\epsilon_{\mathrm{rot}}$ applies per residual
rotation. A complete error budget must account for their multiplicities
and the algorithm's other approximations.

Qubit counts include public registers and peak nested private workspace.
A separate MCX ancilla-pool field must be included when interpreting
strict-lowering workspace. Open programs expose only a known lower bound
and unresolved workspace, since a replacement module can introduce
private locals. Additive gate and access counts, by contrast, multiply
through Repeat and aggregate through calls.

QRAM reads and restricted writes are counted by entry-level logical
resource after call-argument substitution. A query is not assigned the
gate cost of a physical memory implementation.
Finally, the current gate-table and constant-addition constructions are
not assumed optimal. Comparison with a theoretical data-loading recipe
must use the implemented netlist, not substitute an idealized cost
attribute. Broader primitive-scaling sweeps and their commands remain
in the resource-estimation manual.

\bibliographystyle{quantum}
\bibliography{refs}

@inproceedings{green2013quipper,
  author    = {Green, Alexander S. and Lumsdaine, Peter LeFanu and Ross, Neil J. and Selinger, Peter and Valiron, Beno{\^i}t},
  title     = {Quipper: A Scalable Quantum Programming Language},
  booktitle = {Proceedings of the 34th ACM SIGPLAN Conference on Programming Language Design and Implementation (PLDI 2013)},
  year      = {2013},
  pages     = {333--342},
  doi       = {10.1145/2491956.2462177}
}

@inproceedings{svore2018qsharp,
  author    = {Svore, Krysta M. and Geller, Alan and Troyer, Matthias and Azariah, John and Granade, Christopher and Heim, Bettina and Kliuchnikov, Vadym and Mykhailova, Mariia and Paz, Andres and Roetteler, Martin},
  title     = {{Q\#}: Enabling Scalable Quantum Computing and Development with a High-level {DSL}},
  booktitle = {Proceedings of the Real World Domain Specific Languages Workshop 2018 (RWDSL 2018)},
  year      = {2018},
  pages     = {1:1--1:10},
  doi       = {10.1145/3183895.3183901}
}

@article{steiger2018projectq,
  author  = {Steiger, Damian S. and H{\"a}ner, Thomas and Troyer, Matthias},
  title   = {{ProjectQ}: An Open Source Software Framework for Quantum Computing},
  journal = {Quantum},
  year    = {2018},
  volume  = {2},
  pages   = {49},
  doi     = {10.22331/q-2018-01-31-49}
}

@misc{javadiabhari2024qiskit,
  author        = {Javadi-Abhari, Ali and Treinish, Matthew and Krsulich, Kevin and Wood, Christopher J. and Lishman, Jake and Gacon, Julien and Martiel, Simon and Nation, Paul D. and Bishop, Lev S. and Cross, Andrew W. and Johnson, Blake R. and Gambetta, Jay M.},
  title         = {Quantum Computing with {Qiskit}},
  year          = {2024},
  eprint        = {2405.08810},
  archivePrefix = {arXiv},
  primaryClass  = {quant-ph}
}

@misc{cirqdevelopers2020cirq,
  author       = {{Cirq Developers}},
  title        = {Cirq},
  year         = {2020},
  howpublished = {Zenodo},
  doi          = {10.5281/zenodo.4062499},
  note         = {Version 0.8.0; \url{https://github.com/quantumlib/Cirq}}
}

@inproceedings{seidel2023qrisp,
  author    = {Seidel, Raphael and Tcholtchev, Nikolay and Bock, Sebastian and Hauswirth, Manfred},
  title     = {Uncomputation in the {Qrisp} High-Level Quantum Programming Framework},
  booktitle = {Reversible Computation (RC 2023), Lecture Notes in Computer Science},
  year      = {2023},
  pages     = {150--165},
  doi       = {10.1007/978-3-031-38100-3_11}
}

@misc{petric2026qrispbe,
  author        = {Petri{\v{c}}, Matic and Zander, Ren{\'e}},
  title         = {Block-encodings as programming abstractions: The {Eclipse Qrisp BlockEncoding} interface},
  year          = {2026},
  eprint        = {2604.18276},
  archivePrefix = {arXiv},
  primaryClass  = {quant-ph}
}

@article{harrow2009hhl,
  author  = {Harrow, Aram W. and Hassidim, Avinatan and Lloyd, Seth},
  title   = {Quantum Algorithm for Linear Systems of Equations},
  journal = {Physical Review Letters},
  year    = {2009},
  volume  = {103},
  number  = {15},
  pages   = {150502},
  doi     = {10.1103/PhysRevLett.103.150502}
}

@article{costa2022discrete,
  author  = {Costa, Pedro C. S. and An, Dong and Sanders, Yuval R. and Su, Yuan and Babbush, Ryan and Berry, Dominic W.},
  title   = {Optimal Scaling Quantum Linear-Systems Solver via Discrete Adiabatic Theorem},
  journal = {PRX Quantum},
  year    = {2022},
  volume  = {3},
  number  = {4},
  pages   = {040303},
  doi     = {10.1103/PRXQuantum.3.040303}
}

@article{childs2017qls,
  author  = {Childs, Andrew M. and Kothari, Robin and Somma, Rolando D.},
  title   = {Quantum Algorithm for Systems of Linear Equations with Exponentially Improved Dependence on Precision},
  journal = {SIAM Journal on Computing},
  year    = {2017},
  volume  = {46},
  number  = {6},
  pages   = {1920--1950},
  doi     = {10.1137/16M1087072},
  eprint  = {1511.02306},
  archivePrefix = {arXiv}
}

@article{low2019qubitization,
  author  = {Low, Guang Hao and Chuang, Isaac L.},
  title   = {Hamiltonian Simulation by Qubitization},
  journal = {Quantum},
  year    = {2019},
  volume  = {3},
  pages   = {163},
  doi     = {10.22331/q-2019-07-12-163}
}

@inproceedings{gilyen2019qsvt,
  author    = {Gily{\'e}n, Andr{\'a}s and Su, Yuan and Low, Guang Hao and Wiebe, Nathan},
  title     = {Quantum Singular Value Transformation and Beyond: Exponential Improvements for Quantum Hamiltonian Simulation},
  booktitle = {Proceedings of the 51st Annual ACM SIGACT Symposium on Theory of Computing (STOC 2019)},
  year      = {2019},
  pages     = {193--204},
  doi       = {10.1145/3313276.3316366}
}

@article{berry2015taylor,
  author  = {Berry, Dominic W. and Childs, Andrew M. and Cleve, Richard and Kothari, Robin and Somma, Rolando D.},
  title   = {Simulating Hamiltonian Dynamics with a Truncated {Taylor} Series},
  journal = {Physical Review Letters},
  year    = {2015},
  volume  = {114},
  number  = {9},
  pages   = {090502},
  doi     = {10.1103/PhysRevLett.114.090502}
}

@article{martyn2021grand,
  author  = {Martyn, John M. and Rossi, Zane M. and Tan, Andrew K. and Chuang, Isaac L.},
  title   = {Grand Unification of Quantum Algorithms},
  journal = {PRX Quantum},
  year    = {2021},
  volume  = {2},
  number  = {4},
  pages   = {040203},
  doi     = {10.1103/PRXQuantum.2.040203}
}

@article{giovannetti2008qram,
  author  = {Giovannetti, Vittorio and Lloyd, Seth and Maccone, Lorenzo},
  title   = {Quantum Random Access Memory},
  journal = {Physical Review Letters},
  year    = {2008},
  volume  = {100},
  number  = {16},
  pages   = {160501},
  doi     = {10.1103/PhysRevLett.100.160501}
}

@misc{dalzell2025qram,
  author        = {Dalzell, Alexander M. and Gily{\'e}n, Andr{\'a}s and Hann, Connor T. and McArdle, Sam and Salton, Grant and Nguyen, Quynh T. and Kubica, Aleksander and Brand{\~a}o, Fernando G. S. L.},
  title         = {A Distillation-Teleportation Protocol for Fault-Tolerant {QRAM}},
  year          = {2025},
  eprint        = {2505.20265},
  archivePrefix = {arXiv},
  primaryClass  = {quant-ph}
}

@article{liu2021pnas,
  author  = {Liu, Jin-Peng and Kolden, Herman {\O}ie and Krovi, Hari K. and Loureiro, Nuno F. and Trivisa, Konstantina and Childs, Andrew M.},
  title   = {Efficient Quantum Algorithm for Dissipative Nonlinear Differential Equations},
  journal = {Proceedings of the National Academy of Sciences},
  year    = {2021},
  volume  = {118},
  number  = {35},
  pages   = {e2026805118},
  doi     = {10.1073/pnas.2026805118}
}

@article{liu2023carleman,
  author  = {Liu, Jin-Peng and An, Dong and Fang, Di and Wang, Jiasu and Low, Guang Hao and Jordan, Stephen},
  title   = {Efficient Quantum Algorithm for Nonlinear Reaction--Diffusion Equations and Energy Estimation},
  journal = {Communications in Mathematical Physics},
  year    = {2023},
  volume  = {404},
  pages   = {963--1020},
  doi     = {10.1007/s00220-023-04857-9},
  note    = {arXiv:2205.01141}
}

@article{an2026lchs,
  author  = {An, Dong and Childs, Andrew M. and Lin, Lin and Ying, Lexing},
  title   = {Laplace Transform-Based Quantum Eigenvalue Transformation via Linear Combination of {Hamiltonian} Simulation},
  journal = {SIAM Journal on Computing},
  year    = {2026},
  volume  = {55},
  pages   = {376--409},
  doi     = {10.1137/24M1720160}
}

@article{jin2024schrodingerization,
  author  = {Jin, Shi and Liu, Nana and Yu, Yue},
  title   = {Quantum Simulation of Partial Differential Equations via {Schr\"odingerization}},
  journal = {Physical Review Letters},
  year    = {2024},
  volume  = {133},
  pages   = {230602},
  doi     = {10.1103/PhysRevLett.133.230602}
}

@article{hu2024schrocircuits,
  author  = {Hu, Weizhi and Jin, Shi and Liu, Nana and Zhang, Lei},
  title   = {Quantum Circuits for Partial Differential Equations via {Schr\"odingerisation}},
  journal = {Quantum},
  year    = {2024},
  volume  = {8},
  pages   = {1563},
  doi     = {10.22331/q-2024-12-12-1563}
}

@article{wang2026cbmd,
  author  = {Wang, Chao and Liu, Huan-Yu and Xue, Cheng and Zhuang, Xi-Ning and Dou, Menghan and Chen, Zhao-Yun and Guo, Guo-Ping},
  title   = {Quantum Simulation of Non-{Hermitian} Special Functions and Dynamics via Contour-based Matrix Decomposition},
  journal = {Quantum Science and Technology},
  year    = {2026},
  volume  = {11},
  pages   = {035027},
  eprint  = {2511.10267},
  archivePrefix = {arXiv},
  doi     = {10.1088/2058-9565/ae7b7e}
}

@misc{beverland2022assessing,
  author        = {Beverland, Michael E. and Murali, Prakash and Troyer, Matthias and Svore, Krysta M. and Hoefler, Torsten and Kliuchnikov, Vadym and Low, Guang Hao and Soeken, Mathias and Sundaram, Aarthi and Vaschillo, Alexander},
  title         = {Assessing requirements to scale to practical quantum advantage},
  year          = {2022},
  eprint        = {2211.07629},
  archivePrefix = {arXiv},
  primaryClass  = {quant-ph}
}

@misc{harrigan2024qualtran,
  author        = {Harrigan, Matthew P. and Khattar, Tanuj and Yuan, Charles and Peduri, Anurudh and Yosri, Noureldin and Malone, Fionn D. and Babbush, Ryan and Rubin, Nicholas C.},
  title         = {Expressing and {Analyzing} {Quantum} {Algorithms} with {Qualtran}},
  year          = {2024},
  eprint        = {2409.04643},
  archivePrefix = {arXiv},
  primaryClass  = {quant-ph}
}

@article{ross2016optimal,
  author  = {Ross, Neil J. and Selinger, Peter},
  title   = {Optimal ancilla-free {Clifford}+{T} approximation of z-rotations},
  journal = {Quantum Information and Computation},
  volume  = {16},
  number  = {11--12},
  pages   = {901--953},
  year    = {2016},
  doi     = {10.26421/QIC16.11-12-1},
  eprint  = {1403.2975},
  archivePrefix = {arXiv}
}

@misc{qualtran2026interfaces,
  author = {{Qualtran developers}},
  title = {{Qualtran}: Bloq, serialization, and testing interfaces},
  year = {2026},
  howpublished = {\url{https://qualtran.readthedocs.io/en/latest/reference/qualtran.html}},
  note = {Official API documentation, accessed 21 September 2026}
}

@misc{qualtran2026library,
  author = {{Qualtran developers}},
  title = {{Qualtran}: Bloqs Library},
  year = {2026},
  howpublished = {\url{https://qualtran.readthedocs.io/en/latest/bloqs/index.html}},
  note = {Official library documentation, accessed 21 September 2026}
}

@misc{pyliqtr2026features,
  author = {{pyLIQTR developers}},
  title = {{pyLIQTR}: Features},
  year = {2026},
  howpublished = {\url{https://github.com/isi-usc-edu/pyLIQTR/blob/main/docs/Features.md}},
  note = {Official repository documentation, accessed 21 September 2026}
}

@misc{bergholm2018pennylane,
  author        = {Bergholm, Ville and Izaac, Josh and Schuld, Maria and Gogolin, Christian and Ahmed, Shahnawaz and Ajith, Vishnu and others},
  title         = {{PennyLane}: Automatic Differentiation of Hybrid Quantum-Classical Computations},
  year          = {2018},
  eprint        = {1811.04968},
  archivePrefix = {arXiv},
  primaryClass  = {quant-ph}
}

@misc{pennylane2026polytoangles,
  author       = {{PennyLane developers}},
  title        = {{PennyLane}: qml.poly\_to\_angles and {QSVT} documentation},
  year         = {2026},
  howpublished = {\url{https://docs.pennylane.ai/en/stable/code/api/pennylane.poly_to_angles.html}},
  note         = {Official API documentation, accessed 2 October 2026}
}

@misc{catalyst2026compiler,
  author       = {{Catalyst developers}},
  title        = {{Catalyst}: Compiler architecture and {MLIR} dialects},
  year         = {2026},
  howpublished = {\url{https://docs.pennylane.ai/projects/catalyst/en/stable/dev/architecture.html}},
  note         = {Official documentation, accessed 2 October 2026}
}

@misc{obenland2026pyliqtr,
  author       = {Obenland, Kevin and Elenewski, Justin and Morrell, Kaitlyn and Rempfer, Benjamin and Kuklinski, Parker and Neumann, Rylee and Kurlej, Arthur and Rood, Robert and Blue, John and Belarge, Joseph},
  title        = {pyLIQTR},
  year         = {2026},
  howpublished = {Zenodo},
  doi          = {10.5281/zenodo.7221272},
  note         = {QSP/QSVT and block-encoding circuit generation with Clifford+T resource analysis, version 1.4.2; \url{https://github.com/isi-usc-edu/pyLIQTR}}
}

@misc{chuang2021pyqsp,
  author       = {Chuang, Isaac L. and Docter, J. and Martyn, John M. and Rossi, Zane M. and Tan, Andrew K.},
  title        = {pyqsp: Python Quantum Signal Processing},
  year         = {2021},
  howpublished = {\url{https://github.com/ichuang/pyqsp}},
  note         = {Version 0.2.0; QSP phase-angle generation by exact divide-and-conquer and Newton iteration}
}

@misc{unitarylab2025platform,
  author       = {{UnitaryLab}},
  title        = {UnitaryLab Quantum Scientific Computing Platform},
  year         = {2025},
  howpublished = {\url{https://www.unitarylab.com}},
  note         = {Agent-driven studio, proprietary simulator SDK, and MIT-licensed algorithm library (\url{https://github.com/unitarylab/unitarylab_algorithms}); launched November 2025 by the Shanghai Jiao Tong University quantum scientific computing team; accessed September 2026}
}

@misc{classiq2026platform,
  author       = {{Classiq Technologies}},
  title        = {The Classiq Platform and the {Qmod} Quantum Modeling Language},
  year         = {2026},
  howpublished = {\url{https://www.classiq.io}},
  note         = {Constraint-driven circuit synthesis; open example library at \url{https://github.com/Classiq/classiq-library}; accessed September 2026}
}

@book{liao2003beyond,
  author    = {Liao, Shijun},
  title     = {Beyond Perturbation: Introduction to the Homotopy Analysis Method},
  series    = {Modern Mechanics and Mathematics},
  publisher = {Chapman \& Hall/CRC},
  address   = {Boca Raton, FL},
  year      = {2003},
  doi       = {10.1201/9780203491164}
}

@book{liao2012homotopy,
  author    = {Liao, Shijun},
  title     = {Homotopy Analysis Method in Nonlinear Differential Equations},
  publisher = {Springer},
  address   = {Berlin, Heidelberg},
  year      = {2012},
  doi       = {10.1007/978-3-642-25132-0}
}

@article{xue2025qham,
  author  = {Xue, Cheng and Xu, Xiao-Fan and Zhuang, Xi-Ning and Sun, Tai-Ping and Wang, Yun-Jie and Tan, Ming-Yang and Ye, Chuang-Chao and Liu, Huan-Yu and Wu, Yu-Chun and Chen, Zhao-Yun and Guo, Guo-Ping},
  title   = {Quantum Homotopy Analysis Method with Quantum-Compatible Linearization for Nonlinear Partial Differential Equations},
  journal = {Science China Physics, Mechanics \& Astronomy},
  year    = {2025},
  volume  = {68},
  number  = {10},
  pages   = {104702},
  doi     = {10.1007/s11433-024-2584-2},
  eprint  = {2411.06759},
  archivePrefix = {arXiv}
}

@article{chen2022qfvm,
  author  = {Chen, Zhao-Yun and Xue, Cheng and Chen, Si-Ming and Lu, Bing-Han and Wu, Yu-Chun and Ding, Ju-Chun and Huang, Sheng-Hong and Guo, Guo-Ping},
  title   = {Quantum Approach to Accelerate Finite Volume Method on Steady Computational Fluid Dynamics Problems},
  journal = {Quantum Information Processing},
  year    = {2022},
  volume  = {21},
  number  = {4},
  pages   = {137},
  doi     = {10.1007/s11128-022-03478-w},
  eprint  = {2102.03557},
  archivePrefix = {arXiv}
}

@misc{zhuang2025navierstokes,
  author  = {Zhuang, Xi-Ning and Chen, Zhao-Yun and Tan, Ming-Yang and Zhang, Jiaxuan and Ye, Chuang-Chao and Wei, Tian-Hao and Ma, Teng-Yang and Xue, Cheng and Liu, Huan-Yu and Li, Qing-Song and Sun, Tai-Ping and Xu, Xiao-Fan and Wang, Yun-Jie and Wu, Yu-Chun and Guo, Guo-Ping},
  title   = {A Pathway to Practical Quantum Advantage in Solving {Navier--Stokes} Equations},
  year    = {2025},
  eprint  = {2509.08807},
  archivePrefix = {arXiv},
  primaryClass = {quant-ph}
}

@article{chen2024fluid,
  author  = {Chen, Zhao-Yun and Ma, Teng-Yang and Ye, Chuang-Chao and Xu, Liang and Bai, Wen and Zhou, Lei and Tan, Ming-Yang and Zhuang, Xi-Ning and Xu, Xiao-Fan and Wang, Yun-Jie and Sun, Tai-Ping and Chen, Yong and Du, Lei and Guo, Liang-Liang and Zhang, Hai-Feng and Tao, Hao-Ran and Wang, Tian-Le and Yang, Xiao-Yan and Zhao, Ze-An and Wang, Peng and Zhang, Sheng and Zhao, Ren-Ze and Zhang, Chi and Jia, Zhi-Long and Kong, Wei-Cheng and Dou, Meng-Han and Wang, Jun-Chao and Liu, Huan-Yu and Xue, Cheng and Zhang, Peng-Jun-Yi and Huang, Sheng-Hong and Duan, Peng and Wu, Yu-Chun and Guo, Guo-Ping},
  title   = {Enabling Large-Scale and High-Precision Fluid Simulations on Near-Term Quantum Computers},
  journal = {Computer Methods in Applied Mechanics and Engineering},
  year    = {2024},
  volume  = {432},
  pages   = {117428},
  doi     = {10.1016/j.cma.2024.117428},
  eprint  = {2406.06063},
  archivePrefix = {arXiv}
}

@article{yuan2023comparator,
  author  = {Yuan, Yewei and Wang, Chao and Wang, Bei and Chen, Zhao-Yun and Dou, Meng-Han and Wu, Yu-Chun and Guo, Guo-Ping},
  title   = {An Improved {QFT}-Based Quantum Comparator and Extended Modular Arithmetic Using One Ancilla Qubit},
  journal = {New Journal of Physics},
  year    = {2023},
  volume  = {25},
  number  = {10},
  pages   = {103011},
  doi     = {10.1088/1367-2630/acfd52},
  eprint  = {2305.09106},
  archivePrefix = {arXiv}
}

@article{wang2026qramnoise,
  author  = {Wang, Yun-Jie and Sun, Tai-Ping and Zhuang, Xi-Ning and Xu, Xiao-Fan and Liu, Huan-Yu and Xue, Cheng and Wu, Yu-Chun and Chen, Zhao-Yun and Guo, Guo-Ping},
  title   = {Refined Criteria for Quantum Random-Access Memory Error Suppression via an Efficient Large-Scale Simulator},
  journal = {Physical Review Applied},
  year    = {2026},
  volume  = {25},
  number  = {4},
  pages   = {044069},
  doi     = {10.1103/r2pt-dzn8}
}

\end{document}